\documentclass[10pt, twocolumn, journal]{IEEEtran}
\usepackage{cite}
\usepackage{graphicx}
\usepackage{epsf}
\usepackage{subcaption}
\usepackage[labelformat=simple]{subcaption}

\usepackage{amsmath}
\usepackage{comment}
\usepackage{amssymb}
\usepackage{array}
\usepackage{setspace}
\usepackage{amsthm,amssymb}
\usepackage[ruled,lined]{algorithm2e}
\usepackage{multirow}
\usepackage{tabularx}
\usepackage{xfrac}
\usepackage{caption}
\usepackage{breqn}
\usepackage{bbm}
\usepackage{bm}
\usepackage{enumerate}
\usepackage{algorithmic}
\usepackage{dsfont}
\usepackage{float}
\usepackage{diagbox}
\usepackage{mathtools}
\usepackage{adjustbox}
\usepackage{comment}
\usepackage{color}

\graphicspath{{Fig/},{fig/}}

\newcommand{\fig}{Fig.\xspace}

\newcommand{\tabincell}[2]{\begin{tabular}{@{}#1@{}}#2\end{tabular}}
\begin{document}

\title{CRONOS: Colorization and Contrastive Learning for Device-Free NLoS Human Presence Detection using Wi-Fi CSI}

\author{ 
Li-Hsiang Shen,~\IEEEmembership{Member,~IEEE}, 
Chia-Che Hsieh, An-Hung Hsiao, and Kai-Ten Feng,~\IEEEmembership{Senior Member,~IEEE}}


\maketitle

\begin{abstract}
In recent years, the demand for pervasive smart services and applications has increased rapidly. Device-free human detection through sensors or cameras has been widely adopted, but it comes with privacy issues as well as misdetection for motionless people. To address these drawbacks, channel state information (CSI) captured from commercialized Wi-Fi devices provides rich signal features for accurate detection. However, existing systems suffer from inaccurate classification under a non-line-of-sight (NLoS) and stationary scenario, such as when a person is standing still in a room corner. In this work, we propose a system called CRONOS (Colorization and Contrastive Learning Enhanced NLoS Human Presence Detection), which generates dynamic recurrence plots (RPs) and color-coded CSI ratios to distinguish mobile and stationary people from vacancy in a room, respectively. We also incorporate supervised contrastive learning to retrieve substantial representations, where consultation loss is formulated to differentiate the representative distances between dynamic and stationary cases. Furthermore, we propose a self-switched static feature enhanced classifier (S3FEC) to determine the utilization of either RPs or color-coded CSI ratios. Our comprehensive experimental results show that CRONOS outperforms existing systems that either apply machine learning or non-learning based methods, as well as non-CSI based features in open literature. CRONOS achieves the highest human presence detection accuracy in vacancy, mobility, line-of-sight (LoS), and NLoS scenarios.
\end{abstract}

\begin{IEEEkeywords}
	Device-free human presence detection, Wi-Fi, channel state information (CSI), machine learning, deep learning.
\end{IEEEkeywords}

{\let\thefootnote\relax\footnotetext
{Li-Hsiang Shen is with California PATH, Institute of Transportation Studies, University of California at Berkeley (UCB), Berkeley, California, USA. (email: shawngp3@berkeley.edu)}}

{\let\thefootnote\relax\footnotetext
{Chia-Che Hsieh, An-Hung Hsiao, and Kai-Ten Feng are with the Department of Electronics and Electrical Engineering, National Yang Ming Chiao Tung University (NYCU), Hsinchu, Taiwan. (email: chiache.cm09g@nctu.edu.tw,
e.c@nycu.edu.tw, and ktfeng@nycu.edu.tw)}}

\section{Introduction} \label{Intro}

The increasing popularity of smart homes has led to the emergence of many intelligent applications, such as smart lighting, media entertainment, and home energy management systems. Home security applications have also gained significant attention, including elderly care \cite{fallviewer}, intrusion detection \cite{intrusion}, and disaster prevention. For these applications, indoor human presence detection is a necessary and core technology, particularly in device-free scenarios where electronic devices are not required on the targeted individuals. Infrared sensors \cite{infrared} and camera-based methods \cite{camera} are currently the most common device-free human presence detection technologies \cite{acm}. Infrared sensors are advantageous since they are low cost and energy-efficient. However, they only work when a person is moving within line-of-sight (LoS) and are ineffective for non-line-of-sight (NLoS) or motionless scenarios. While camera-based methods can provide more comprehensive image information, they are impeded by privacy concerns and blindspot issues. Wi-Fi signals, on the other hand, can pass through walls and do not raise privacy concerns, making them a prevalent research topic in human presence detection.

The Wi-Fi signal channel state information (CSI) can be estimated through channel sounding. Initially designed to represent channel characteristics and to generate an appropriate equalizer compensating for wireless channel effects, CSI is now widely utilized in wireless sensing tasks since human activities also affect the wireless channel. CSI provides finer-grained information compared to received signal strength (RSS) and can be decomposed into amplitude and phase information. Recent studies proposed several wireless sensing tasks utilizing CSI amplitude or phase information, such as indoor localization \cite{loc_amp,loc_phase}, human counting \cite{hc1,hc2}, vital sign monitoring during sleep \cite{vs_amp}, and respiration and heartbeat monitoring \cite{phasebeat}. However, these studies only adopted either the amplitude or phase information of CSI, which leads to incomplete usage of CSI information weakening its detection ability. To improve the accuracy and expand the detection area, some studies, such as \cite{csi_image} and \cite{farsense}, utilized both amplitude and phase information to achieve their tasks. The CSI ratio, which combines both amplitude and phase information in the form of a complex number, is considered suitable for human presence detection since it can provide stronger detection capabilities. Converting the CSI ratio into a complex plane image is a novel design to represent the complete features of the CSI ratio, which is suitable for indoor human presence detection as it can cover the entire indoor space and avoid detection blindspots.

Human activities have a time component, which means that CSI data exhibits corresponding temporal characteristics. Therefore, in many wireless sensing studies, the time series data of CSI is used for classification \cite{ts1}, \cite{ts2}. Moreover, time series of CSI can be converted into images with temporal features, such as recurrence plots (RPs) \cite{rp}. RPs are diagrams that can analyze the recurrence in a time series and identify whether the signal is periodic or chaotic during a specific time period. RPs have been widely utilized in qualitative analysis by calculating some parameters in various fields. For example, in \cite{rp_sea}, RPs are constructed from radar sea clutter data to classify sea surface floating and sea clutter, and the recurrence rate is computed from RP to aid classification. In \cite{rp_gastric}, RPs are generated from gastric slow waves to calculate the recurrence rate and diagonal line entropy for gastrointestinal field analysis. In CSI-based wireless sensing tasks, the authors of \cite{wi_sleep} and \cite{rp_respiration} aim to implement a respiration detection system. They generate RPs from all subcarrier values of CSI data and adopt the periodicity of RPs to perform subcarrier selection. With the increasing popularity of RPs for representing temporal features in various fields, we propose to use RPs in human presence detection to distinguish between an empty room and the presence of a moving person in environments with time-changing human behaviors. Moreover, the use of RPs can enable the development of powerful image classification algorithms to enhance accuracy in human presence detection.

After converting CSI into images, a learning method is required to classify these images, and contrastive learning techniques have become increasingly popular in recent years. Initially, SimCLR \cite{simclr} proposed a simple framework for contrastive learning on visual representation, achieving top-1 accuracy on ImageNet using a self-supervised learning algorithm for image recognition. Since then, many papers have adopted the architecture of self-supervised contrastive learning \cite{sscl1,sscl2,sscl3}. However, self-supervised learning does not utilize complete label information. Therefore, SupCon \cite{supcon} proposes supervised contrastive learning that can incorporate label information. Although the cross-entropy loss is widely employed for supervised classification problems, several papers have shown that it lacks robustness to noisy labels \cite{noisy_label1, noisy_label2} and can result in poor margins \cite{poor_margins1, poor_margins2}. SupCon confirms that their supervised contrastive loss outperforms the cross-entropy loss on supervised learning. Some wireless sensing applications employ self-supervised or supervised contrastive learning architectures to implement their work. For example, STF-CSL \cite{STF-CSL} combines self-supervised contrastive learning with time-domain and frequency-domain data augmentation for human activity recognition tasks. Additionally, the paper \cite{loc_scl} uses the architecture of supervised contrastive learning to perform fingerprint-based positioning on real-world outdoor CSI data. Contrastive learning has been successful in many fields, particularly in computer vision. For human presence detection, contrastive learning is a powerful classification algorithm for converting CSI into images for classification. However, labeled data is required for presence detection using the learning method, and supervised contrastive learning can be more effective than the cross-entropy loss in addressing the problem of noisy labels.

Recently, several studies have proposed the use of CSI signals for human presence detection. For instance, \cite{pd_svm} has adopted density-based spatial clustering to reduce noise influence on CSI signals and utilized support vector machines (SVM) to determine whether there is a standing or walking person in a room. In \cite{f_lstm}, the authors filtered CSI data using Butterworth and moving average filters and then utilized long short-term memory (LSTM) to classify whether there is someone in the indoor space. In \cite{p_cnn}, CSI amplitude and phase information were transformed into images, and two parallel convolutional neural networks (CNNs) were employed to extract the amplitude and phase features for classifying the empty and walking human cases. Similarly, in \cite{yuming}, a convolutional denoising autoencoder was proposed to reduce the dimensionality of CSI data, which was then inputted into a neural network for multiple spot presence detection. In \cite{fangyu}, dynamic and spatial domain features were extracted in data preprocessing, and a conditional recurrent neural network was designed for multi-room human presence detection. However, none of these studies addressed the issue of NLoS standing persons. Since a stationary person in the corner may not cause noticeable disturbance to the wireless channel, the signal characteristics of the CSI data can be similar to that of an empty room. Thus, it is necessary to address the NLoS static problem to achieve accurate human presence detection in realistic circumstances.

As mentioned in the previous paragraphs, there have been several effective designs for CSI-based human presence detection, such as CSI ratio, RPs, and contrastive learning. However, this work is the first to incorporate these techniques into a comprehensive human presence detection system. Additionally, previous studies that used CSI signals for indoor human presence detection did not address the issue of distinguishing an empty room from a stationary person in NLoS corners. Therefore, we propose the colorization and contrastive learning enhanced NLoS human presence detection (CRONOS) system. The colorization algorithm for the CSI ratio image is designed to solve the NLoS static problem, and RPs are leveraged to determine whether a person is walking or not. Finally, supervised contrastive learning is employed to learn the representations and achieve human presence detection that can detect a stationary person in NLoS corners. The main contributions of this work are listed as follows.

\begin{itemize}
	\item The proposed CRONOS system can detect indoor human presence by classifying an empty room, an NLoS standing person, an LoS standing person, and a walking person. We are the first to design a system that can detect stationary people in NLoS corners, whereas existing methods only deal with moving or static people in locations that significantly affect the CSI signal, such as LoS.

	\item The CRONOS system comprises feature-encompassed image generation (FEIG) and three-stage supervised contrastive learning. FEIG generates dynamic and static feature images. Dynamic feature (DF) images are RPs generated from the CSI amplitude difference between antennas and can distinguish dynamic and static situations. Coloring CSI ratio images generated by the colorization algorithm are static feature (SF) images that help separate static cases, addressing the NLoS static problem.

	\item After transforming CSI data into images in FEIG, we propose three-stage supervised contrastive learning to classify these images into four cases of human presence detection. Stage 1 uses supervised contrastive loss to train the representations of RPs. Stage 2 employs our designed consultation loss to learn the representations of coloring CSI ratio images, which are more capable of resolving the NLoS static problem. In the last stage, we design the self-switched static feature enhanced classifier (S3FEC) to determine utilization of either representations of RP or CSI ratio.

	\item We evaluate the performance of our CRONOS system in two real-world scenarios. Our system achieves the highest F1-score in both the NLoS standing case and overall performance compared to other existing methods that did not have a specific design for identifying NLoS stationary person, demonstrating that our system can solve the NLoS static problem. The ablation study confirmed that each component design in our system significantly contributes to its performance.
\end{itemize}

The rest of this paper is organized as follows. Section \ref{CH_SA} describes the system architecture of the problem and the preliminary of CSI. The proposed CRONOS is explained in Section \ref{CH_Propose}. Section \ref{CH_Exp} provides performance evaluation, whereas the conclusions are drawn in Section \ref{CH_conclusion}.

\section{System Architecture and Preliminary Observations} \label{CH_SA}

\subsection{System Architecture}
\label{sys_arc}

\begin{figure}
\centering
\includegraphics[width=3.5in]{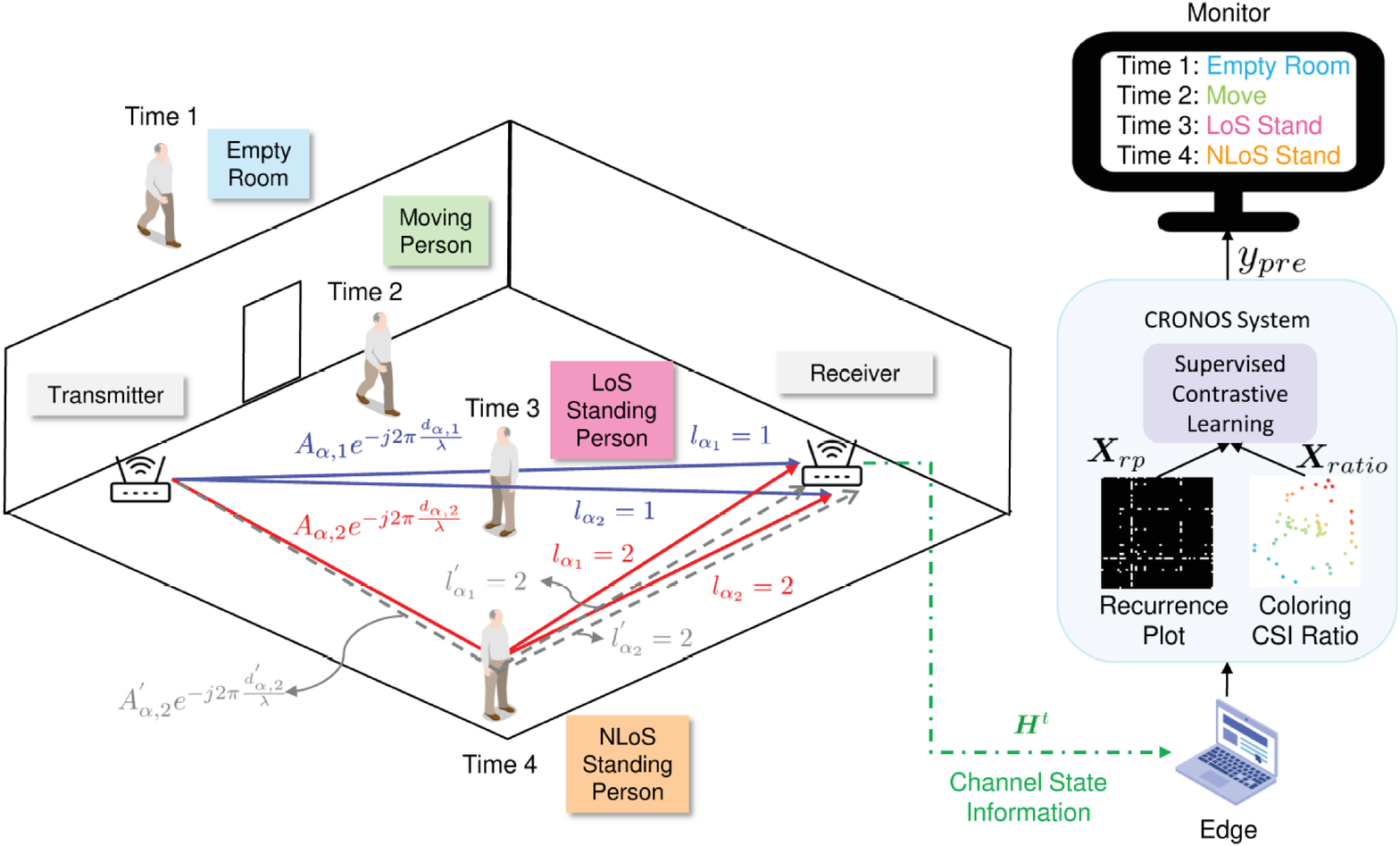}
\caption{System architecture of indoor human presence detection using Wi-Fi CSI.}
\label{fig:system}
\end{figure}

Our objective is to develop a device-free human presence detection system that can accurately detect whether there is a standing person in an indoor environment, specifically in NLoS scenarios. Our system not only distinguishes between an empty room and a room with human presence but also categorizes three different human behaviors in the room. The system architecture is shown in \fig\ref{fig:system}, where two commercially available Wi-Fi access points (APs) are deployed and served as the transmitter and receiver. The receiver estimates CSI by collecting data packets transmitted by the transmitter. Different human presence scenarios result in various multipath behaviors, which trigger the algorithm design for presence detection based on CSI signals. In this context, we consider four cases as follows:
\begin{itemize}
    \item Case 1 (\textbf{Empty}): An empty room without any person in it.
    \item Case 2 (\textbf{Static NLoS}): A person standing still in the NLoS area.
    \item Case 3 (\textbf{Static LoS}): A person standing in the LoS path.
    \item Case 4 (\textbf{Dynamic}): A person walking around the room.
\end{itemize}
We have chosen to focus on these cases of human presence since they cover most situations in daily life\textsuperscript{\ref{note0}}\footnotetext[1]{Detecting the presence of breathing can be another potential solution for human presence detection. However, there exist some limitations in breath detection as demonstrated in \cite{breath1, breath2}. The transceivers should be placed close to the person to be tested in order to detect the signals fluctuation induced by periodic chest movement. Tracking can be another potential solution for detecting an empty room and human walking. However, it requires a more complex system and deep learning design. Moreover, efficient collection and labeling for a large amount of data as well as the frequent deep learning model update are further potential designs. \label{note0}}. The receiver sends the collected CSI data to edge nodes for storage and processing. The system will then predict the detection results which are displayed on a monitor for examination.

\subsection{Channel State Information} 
\label{csi}

We use APs that operate in the multi-input multi-output (MIMO) and orthogonal frequency division multiplexing (OFDM) mechanisms of the IEEE 802.11n protocol. MIMO-OFDM provides tolerance to multipath delay, high channel diversity, and improved link quality, resulting in diversified CSI information. In the MIMO-OFDM system, the CSI data in one packet consists of multiple complex values that offer fine-grained information, and the quantity depends on the number of transmission pairs and OFDM subcarriers. These CSI signals represent the channel property influenced by various phenomena such as reflection, refraction, diffraction, and scattering. Therefore, different furniture placement or human behavior can alter the CSI signal features in the indoor environment. We define the received CSI matrix at time $t$ as
\begin{equation} \label{HT}
    \bm{H}^t = \begin{bmatrix}
                \bm{h}^t_{1,1} & \dots & \bm{h}^t_{1,n} & \dots & \bm{h}^t_{1,N}\\
                \vdots & \ddots & \vdots & \dots & \vdots\\
                \bm{h}^t_{m,1} & \dots & \bm{h}^t_{m,n} & \dots & \bm{h}^t_{m,N}\\
                \vdots & \ddots & \vdots & \dots & \vdots\\
                \bm{h}^t_{M,1} & \dots & \bm{h}^t_{M,n} & \dots & \bm{h}^t_{M,N}
               \end{bmatrix},\quad
\end{equation}
where $M$ and $N$ indicates the number of transmitter and receiver antennas, respectively. In $\eqref{HT}$, each element represents a vector for the $m$-th transmit antenna and the $n$-th receiving antenna, i.e., the CSI data of a transmission pair, which is defined as
\begin{equation}
    \bm{h}^t_{m,n} = \left[h^t_{m,n,1}, h^t_{m,n,2}, \dots, h^t_{m,n,k}, \dots, h^t_{m,n,K} \right],
\end{equation}
where $k \in \left[1, \dots, K\right]$ is the subcarrier index with $K$ as the number of available subcarriers. Each element of $\bm{h}^t_{m,n}$ is a superposition of all signal paths, which characterizes the multipath propagation in respective subcarriers. As a result, the CSI value can be modeled as
\begin{align}
    h^t_{m,n,k} &= e^{-j\phi^t_{m,n,k}}\sum_{l=1}^{L_{m,n}}A^t_{m,n,l}e^{-j2\pi\frac{d^t_{m,n,l}}{\lambda_k}} \notag\\
    &= |h^t_{m,n,k}|e^{j(\angle{h^t_{m,n,k}})},
\label{eq:csi_value}
\end{align}
where $\phi^t_{m,n,k}$ is a random phase offset, and $L_{m,n}$ is the number of propagation paths that arrive at the $n$-th receiver antenna from the $m$-th transmitter antenna. $A^t_{m,n,l}$ and $d^t_{m,n,l}$ indicate the complex attenuation and propagation length of the $l$-th path, respectively. $\lambda_k$ is the wavelength of the $k$-th subcarrier. Moreover, $|h^t_{m,n,k}|$ and $\angle{h^t_{m,n,k}}$ denote the amplitude and phase of the transmission pair of the $m$-th transmitter and $n$-th receiver antenna at the $k$-th subcarrier, respectively. In CSI sensing research, both amplitude and phase information can be crucial components as they are affected by human activity, causing amplitude attenuation and phase shift. However, due to the lack of perfect time or frequency synchronization in Wi-Fi APs, there exists a random phase offset $\phi^t_{m,n,k}$ which makes it difficult to use raw CSI phase information for sensing tasks. Despite this issue, we still aim to incorporate the benefits of CSI phase data in our human presence detection system. To overcome the problem of raw phase, we use the ratio between CSI values from different antennas instead. The explanation and derivation of this approach will be presented in the next section.

\begin{figure}
\centering
  \begin{subfigure}[b]{0.45\textwidth}
    \includegraphics[width=\textwidth]{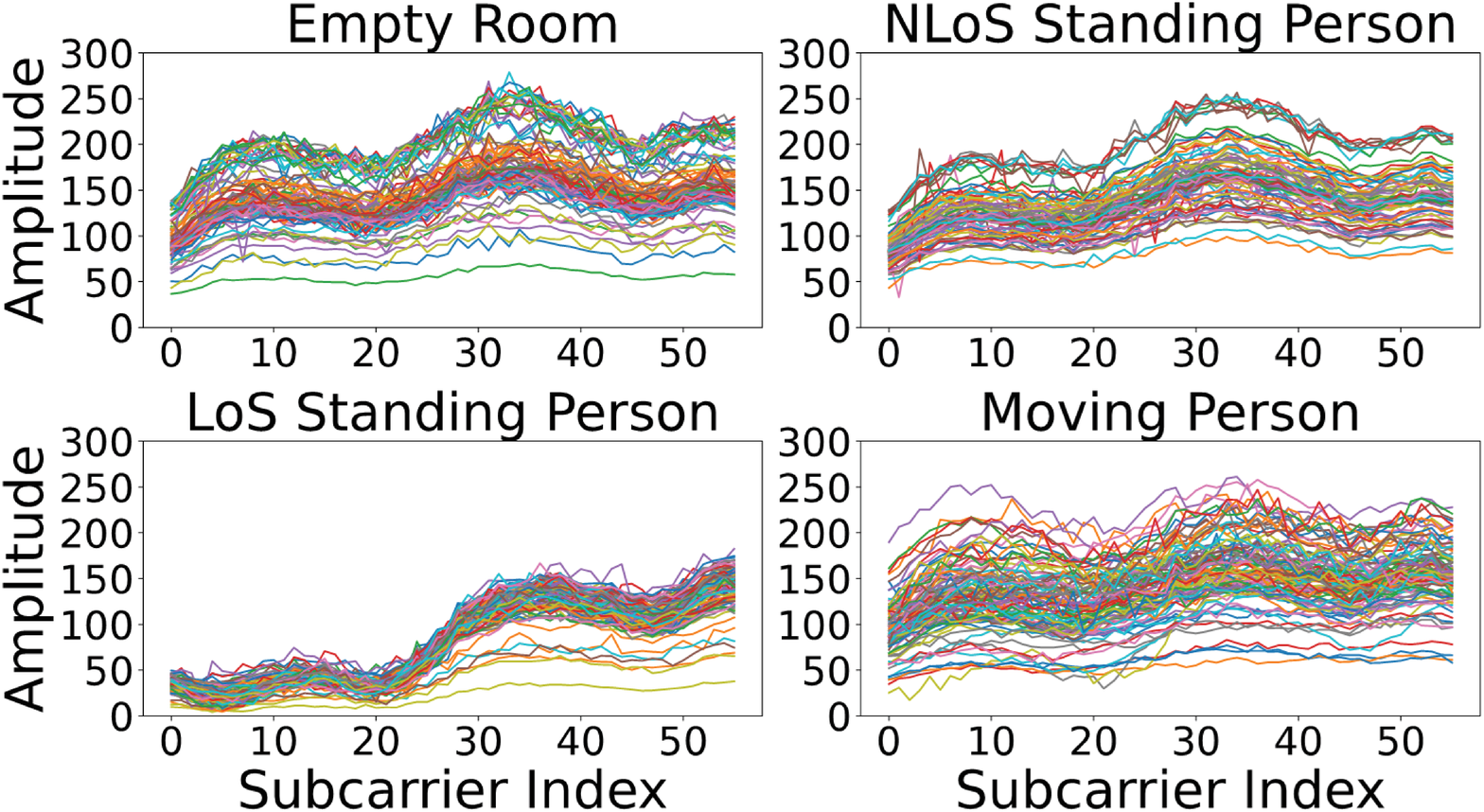}
    \caption{\footnotesize}
    \label{fig:csi_amp}
  \end{subfigure}
  \begin{subfigure}[b]{0.45\textwidth}
    \includegraphics[width=\textwidth]{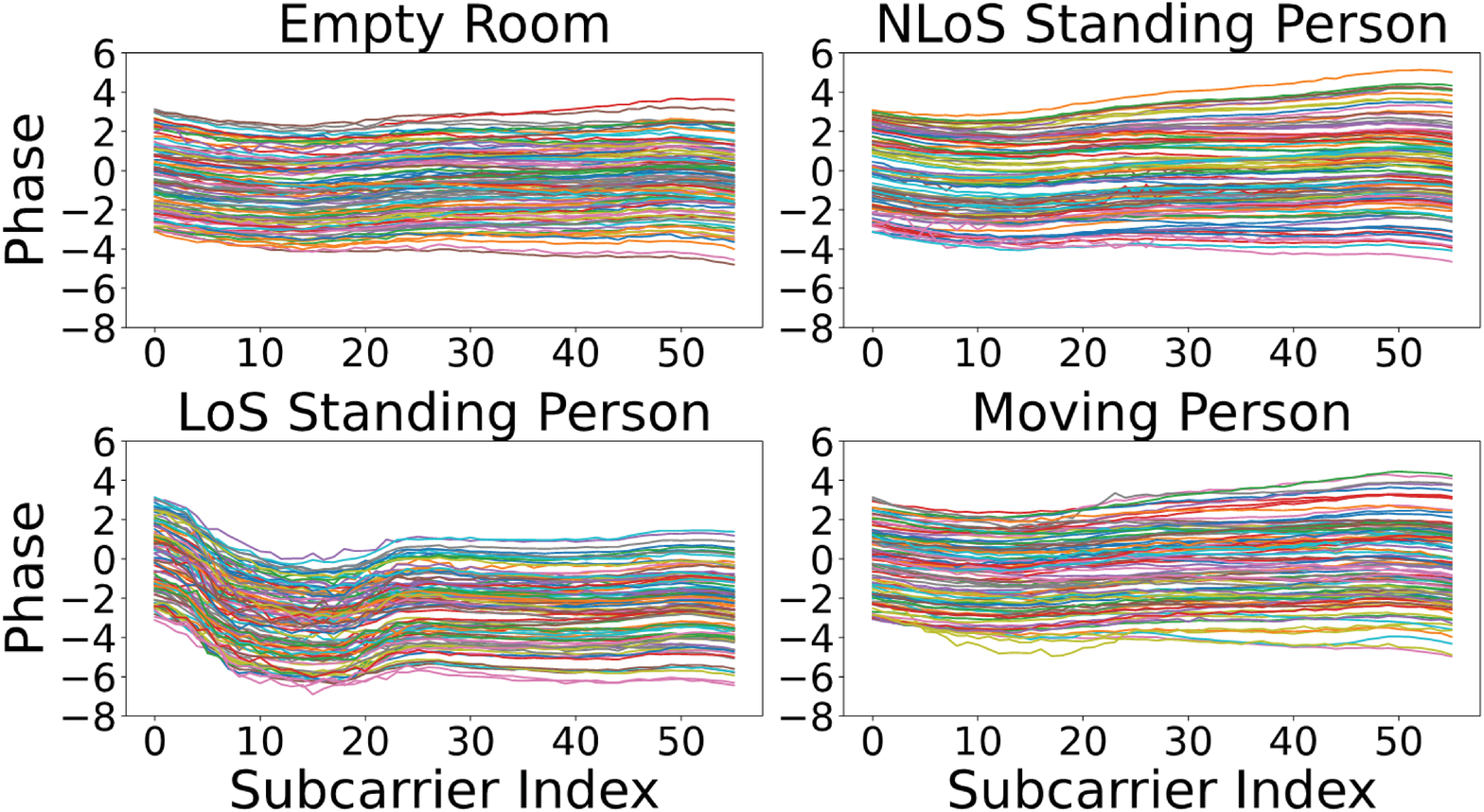}
    \caption{\footnotesize}
    \label{fig:csi_phase}
  \end{subfigure}
\caption{The measured CSI data in each subcarrier for four cases, including (a) amplitude and (b) phase.}
\label{fig:csi_data}
\end{figure}

\subsection{CSI Observation of Human Presence}
\label{pre_exp}
Since rich multipath signals reach the receiver in the indoor scenario, the CSI data can effectively reflect human activity and environmental change. Hence, the primary goal of our system is to differentiate the four cases by CSI signals. However, using CSI for classifying these four cases is not simple, especially in the NLoS standing case. For this reason, we conduct preliminary experiments to observe CSI amplitude and phase information under different scenarios, assisting us to comprehend the ambiguity of NLoS and gain insights for formulating effective approaches. The following results are exemplified observations from a transmission pair.

\subsubsection{NLoS Static Problem}
\label{nlos_sp}

\begin{figure}
\centering
  \begin{subfigure}[b]{0.45\textwidth}
    \includegraphics[width=\textwidth]{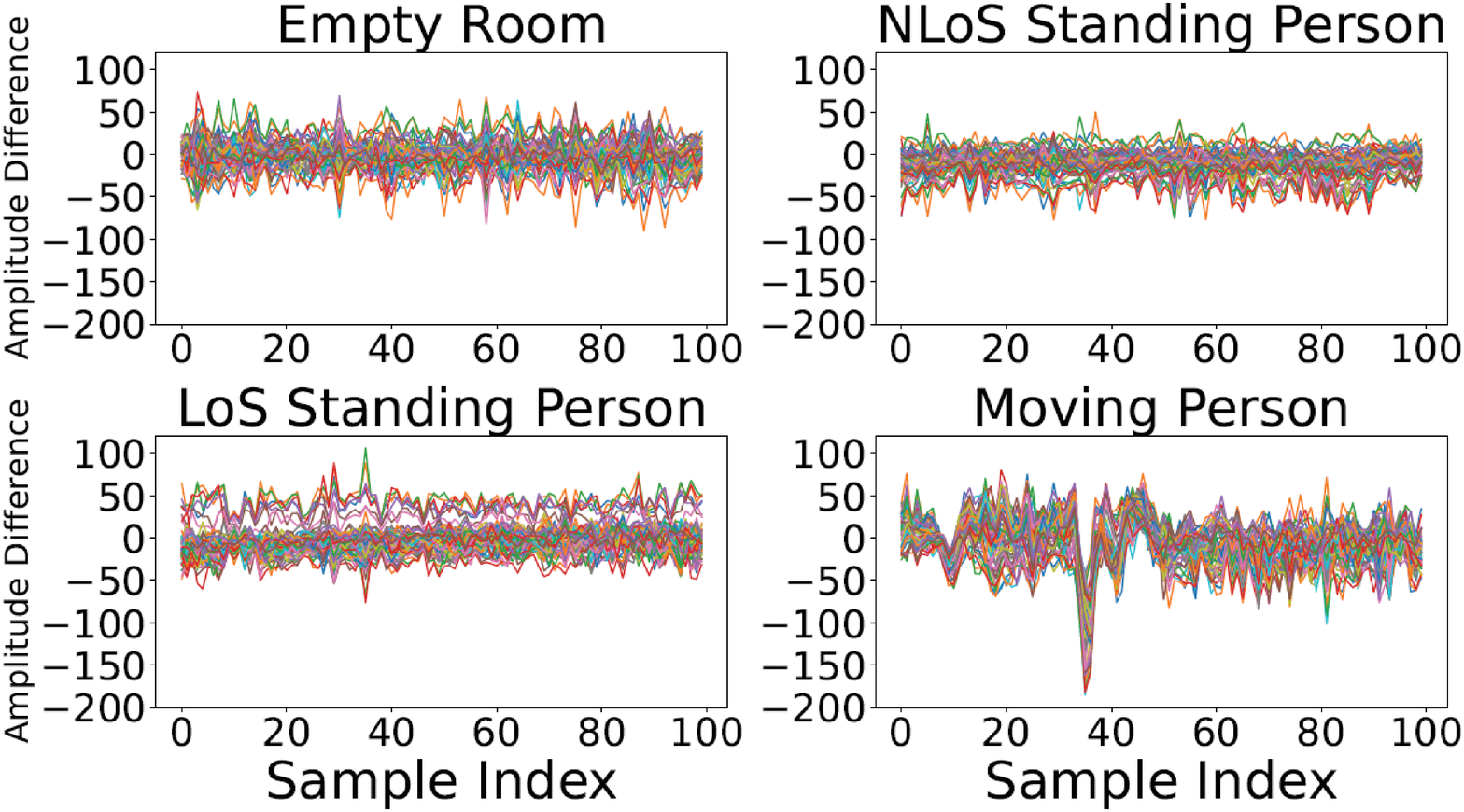}
    \caption{\footnotesize}
    \label{fig:csi_amp_diff}
  \end{subfigure}
  \begin{subfigure}[b]{0.45\textwidth}
    \includegraphics[width=\textwidth]{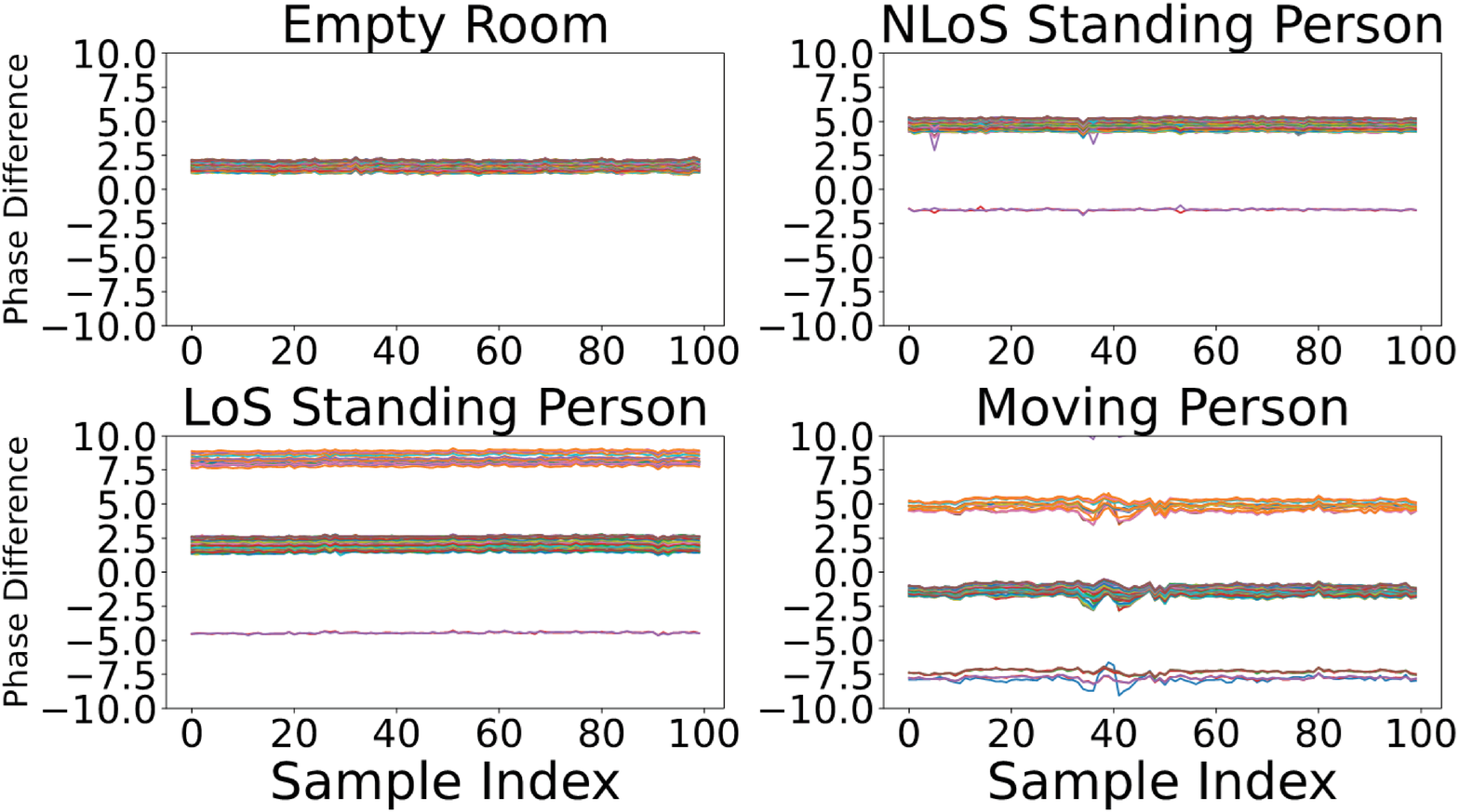}
    \caption{\footnotesize}
    \label{fig:csi_phase_diff}
  \end{subfigure}
\caption{The (a) amplitude difference and (b) phase difference versus time samples for four cases.}
\label{fig:csi_data2}
\end{figure}

As shown in \fig\ref{fig:csi_amp}, we compare the CSI amplitude information of four cases. In each case, different curves represent 100 sampling data collected from continuous packets over time. We observe that the amplitude patterns in the empty room and NLoS standing case are quite similar. This is because those attenuated signal paths in the two situations do not significantly impact the overall CSI amplitude value. Moreover, both cases represent static conditions, which their CSI amplitudes exhibit no significant fluctuations over time by observing different curves in each plot. Therefore, it is challenging to differentiate between these two cases using only CSI amplitude information. On the other hand, the shapes of CSI amplitude in the LoS standing case are vastly different from those in the empty room and NLoS case due to the blocked LoS path. Additionally, the CSI amplitude of human movement will fluctuate over time because the walking person causes substantial changes in the wireless channel. These influences enable us to easily distinguish the LoS standing and moving scenarios from other cases. The above observations and identical explanations can also be found in the amplitude difference in Fig. \ref{fig:csi_amp_diff}.

Next, we have collected the CSI phase information of the four cases using the same samples as the amplitude data. The raw CSI phase was unwrapped along the time axis. We observe from the Fig. \ref{fig:csi_phase} that CSI phase is substantially fluctuating owing to contamination with the random phase offset \cite{arx1} as mentioned in $\eqref{eq:csi_value}$. This renders the raw CSI phase infeasible for sensing tasks. Nevertheless, such situation can be mitigated by employing phase difference in Fig. \ref{fig:csi_phase_diff}, as all antennas share a common oscillator. From Figs. \ref{fig:csi_phase} and \ref{fig:csi_phase_diff}, distinguishing between the majority of cases using solely phase or phase difference information is observed to be challenging. However, exceptions include the LoS case in the phase plot and the moving case in the figure of phase difference. Furthermore,  establishing precise thresholds to differentiate the averaged phase difference is critical, as they rely on the unwrap operation, various environmental conditions and diverse device characteristics. Slight peaks for the NLoS static case in Fig. \ref{fig:csi_phase_diff} are observed in only few samples, which can be attributed to multipath effects caused by NLoS objects. In summary, utilizing either CSI amplitude/phase or their differences alone imposes a challenge in accurately classifying all scenarios.

\subsubsection{Dynamic Feature for Walking Person}
\label{DF}


We also conduct experiments to observe the CSI amplitude difference between two receiving antennas among the four cases versus time samples as depicted in \fig\ref{fig:csi_amp_diff}. We can infer that the moving case exhibits a larger fluctuation, whilst the values in the other cases remain relatively stable. This is because the empty room, NLoS standing person, and LoS standing person perform the static scenarios. The signal paths and attenuation between the transmitter and the two receiving antennas do not vary over time, resulting in a consistent amplitude difference in these cases. However, in the moving case, the person's motion introduces different impacts on the signals at different timestamps. For example, if the moving person blocks the LoS path of the first antenna at time instant $t_1$ but not at $t_2$, the CSI values will show significant differences between the two antennas at these two timestamps. Thus, the amplitude difference values will oscillate noticeably. Moreover, we can observe from Fig. \ref{fig:csi_phase_diff} that the phase difference also provides informative feature when detecting a moving person, whilst the others perform similar curves. To elaborate a little further, the commercial AP device should be perfectly calibrated to extract the precise phase information, which however takes time and is considered impractical in real applications. Once we adopt the infeasible CSI phase part, it may degrade the detection precision. Neither adopting phase difference nor using raw phase data can help detect amongst static cases. Based on this characteristic, we can develop an approach to detect whether a person is dynamically moving or stationary in the room.

\section{Proposed CRONOS System} \label{CH_Propose}

\begin{figure*}
\centering
\includegraphics[width=0.8\linewidth]{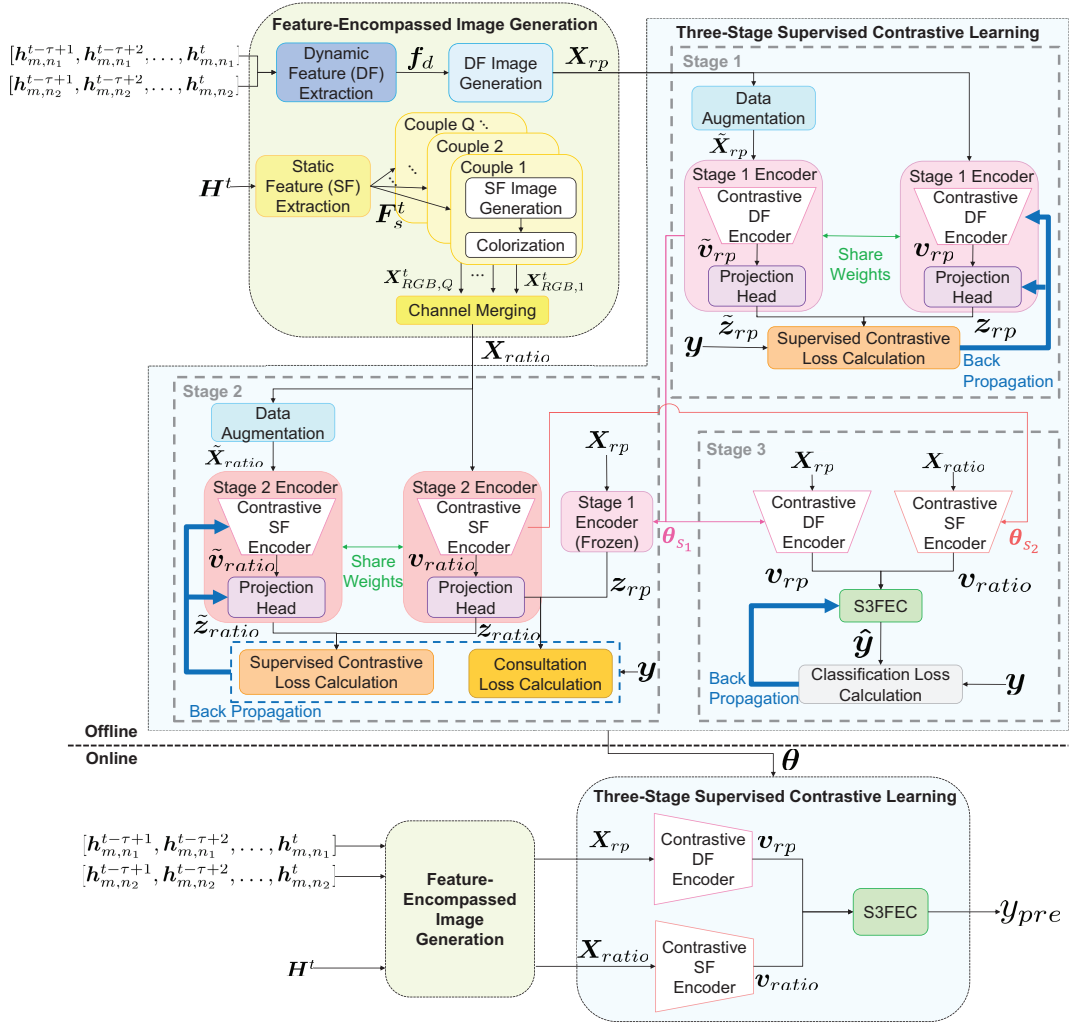}
\caption{Block diagram of the proposed CRONOS system.}
\label{fig:block_diagram}
\end{figure*}

In the previous section, we have observed that relying solely on CSI amplitude or phase information is insufficient to resolve the NLoS static problem. Moreover, we can employ the CSI amplitude difference between antennas to detect whether a person is moving or not. In this section, we describe our approach to address the NLoS static problem and classify the presence of a person into three behaviors. The block diagram of our CRONOS system for presence detection is shown in \fig\ref{fig:block_diagram}, which consists of offline and online phases. The offline phase has two parts, including FEIG and three-stage supervised contrastive learning. In the FEIG process, we convert the CSI data into images to turn our presence detection problem into an image recognition problem. The DF extraction function extracts DF from a CSI time series of two transmission pairs, which is then converted into an image as the input to separate the dynamic case from static ones. The SF extraction is conducted at the final time slot of the CSI sequence, and the resulting SF couples generate a coloring image using the colorization algorithm. These images are merged as the input for distinguishing the three static cases. 

Furthermore, three-stage supervised contrastive learning trains the contrastive encoder for the DF in stage 1, where the goal is to obtain dynamic representations for the four cases. Stage 2 focuses on the contrastive encoder for the SF and combines the consultation loss to assist the model in getting better static representations. In stage 3, the DF and SF encoders obtained from stages 1 and 2 are frozen and associated with the S3FEC to perform the feature map. Afterwards, the online models are updated by computing the designed loss function as well as back-propagation mechanism. During the online phase, we convert the real-time CSI data into images using FEIG. Our system predicts the real-time results of human presence detection by inputting the image data into the well-trained model.

\subsection{Feature-Encompassed Image Generation (FEIG)}
\label{FEIG}
In this subsection, we will provide a detailed description of each block in FEIG. As mentioned earlier, our system converts the CSI data into images through both the DF and SF image generation blocks. Additionally, we utilize colorization and channel merging to enhance the SF. Therefore, FEIG is a data preprocessing designed to generate images that capture both dynamic and static features of CSI.

\subsubsection{Dynamic Feature Extraction and Recurrence Plot}
As shown in Fig. \ref{fig:csi_amp_diff}, the amplitude difference in the human walking in case 4 has acute variations along the time axis, while the other three static cases 1 to 3 are relatively stable. We use the amplitude difference as the DF in our work to distinguish between stationary and human motion situations. The amplitude difference of two transmission pairs from the $m$-th transmitter antenna at time $t$ and $k$-th subcarrier can be expressed as
\begin{equation}
    d^t_{m,n_d,k} = \left|h^t_{m,n_1,k}|-|h^t_{m,n_2,k}\right|,
\end{equation}
where $n_1,n_2 \in [1,N]$ and $n_1 \ne n_2$. Note that $n_d = (n_1, n_2)$ represents the pair of two receiving antennas $n_1$ and $n_2$, whereas the subscript $d$ means the difference. In addition, to make the characteristics for each case more prominent and easier to analyze, we calculate the average of all subcarrier values, which can be given by
\begin{equation}
    \bar{d}^t_{m,n_d} = \frac{1}{K}\sum_{k=1}^{K}d^t_{m,n_d,k}.
\end{equation}
Next, we need to produce a time window for our DF since these features are only observed within a certain time period. The window size determines how many past samples are considered for feature extraction, which can be represented as
\begin{equation}
    \bm{f}_d = [\bar{d}^{t-\tau+1}_{m,n_d}, \bar{d}^{t-\tau+2}_{m,n_d}, \dots, \bar{d}^{t}_{m,n_d}],
\label{eq:fd}
\end{equation}
where the $\tau$ is the window size. 

The next step is to convert the DF from numerical values into DF images to enhance the distinction between static and dynamic cases. For this purpose, we employ RPs \cite{rp} as the DF image. RPs are graphs that can determine whether a statistical value or system returns to its original state at a certain time. If the values are stochastic, the points in RPs are more uniform. If the values possess periodic property, RPs reveal regular structures. Our CRONOS system employs RPs in image form as our classification input, which can be obtained from DF calculated in $\eqref{eq:fd}$. The RP matrix is represented as
\begin{equation}
    \bm{X}_{rp} = \begin{bmatrix}
                x^{t,t-\tau+1}_{rp} & x^{t,t-\tau+2}_{rp} & \dots & x^{t,t}_{rp}\\
                x^{t-1,t-\tau+1}_{rp} & x^{t-1,t-\tau+2}_{rp} & \dots & x^{t-1,t}_{rp}\\
                \vdots & \vdots & \ddots & \vdots\\
                x^{t-\tau+1,t-\tau+1}_{rp} & x^{t-\tau+1,t-\tau+2}_{rp} & \dots & x^{t-\tau+1,t}_{rp}
               \end{bmatrix},
\end{equation}
where each element is denoted as
\begin{equation}
    x^{t_1,t_2}_{rp} = 
    \begin{cases}
        1, & \mbox{if }|\bar{d}^{t_1}_{m,n_d} - \bar{d}^{t_2}_{m,n_d}| \le \gamma, \\
        0, & \mbox{otherwise},
    \end{cases}
\label{eq:rp_element}
\end{equation}
with $t_1, t_2 \in \left[ t-\tau+1, t \right]$ and $\gamma$ as the threshold of RPs. The threshold $\gamma$ is determined based on the average absolute difference $\bar{d}^{t}_{m,n_d}$ in a specific time interval of the empty room case. If the absolute value between two consecutive CSI amplitude difference is less than or equal to $\gamma$, the corresponding pixel value in the image is set to 1, indicating black. Otherwise, the pixel is set to 0, indicating white. In order to obtain $\gamma$, we define the following set as
\begin{equation}
    \bm{D}_e = \{ \left|\bar{d}^{t_1}_{m,n_d}(e) - \bar{d}^{t_2}_{m,n_d}(e)\right| | \forall t_1, t_2 \in [t-\tau_{\gamma}+1, t]\},
\label{eq:can_set}
\end{equation}
where $e$ represents the first case of an empty room, and $\tau_{\gamma}$ is the window size for the considered time interval. We select $\gamma = |\bar{d}^{t_1}_{m,n_d}(e) - \bar{d}^{t_2}_{m,n_d}(e)|$ in $\bm{D}_e$ that makes the cumulative distribution function (CDF) to be higher than 0.5, such that the average value of the amplitude difference is less likely to exceed $\gamma$. As a result, most of the elements in the RPs will be shown in black in the human absence case. The NLoS and LoS cases have a similar average value of amplitude difference, which leads to an equivalent result. Since the dynamic case has a perturbation in amplitude difference, it is reasonable to exceed the value of $\gamma$, which will result in white blocks in the RP and can therefore effectively distinguish between static and dynamic cases.

\begin{figure}
  \centering
  \begin{subfigure}[b]{0.25\textwidth}
    \includegraphics[width=\textwidth]{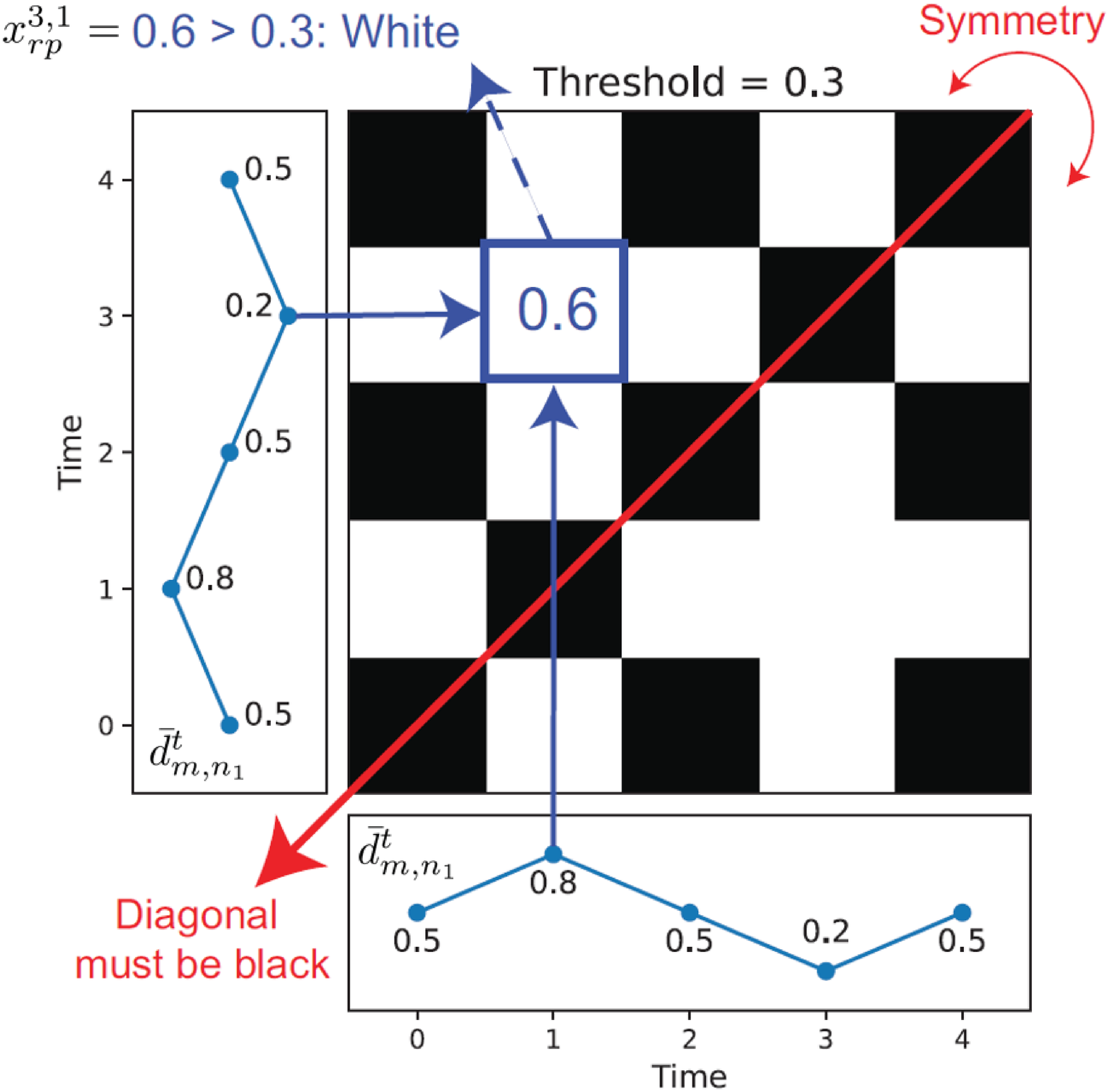}
    \caption{\footnotesize}
    \label{fig:rp_sd_a}
  \end{subfigure}
  \begin{subfigure}[b]{0.23\textwidth}
    \includegraphics[width=\textwidth]{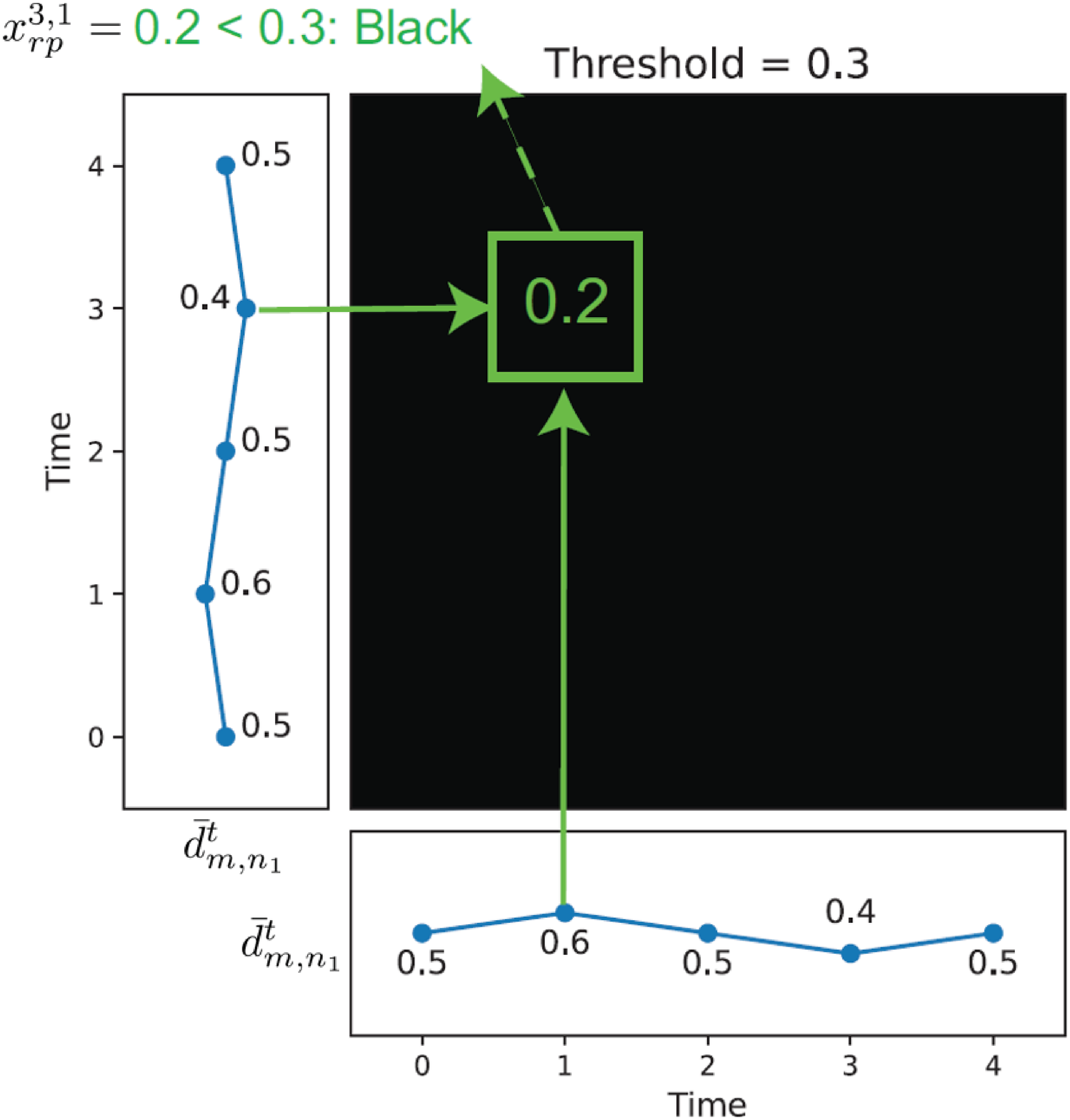}
    \caption{\footnotesize}
    \label{fig:rp_sd_b}
  \end{subfigure}
\caption{The schematic diagram of recurrence plots with (a) larger fluctuation and (b) smaller fluctuation.}
\label{fig:rp_sd}
\end{figure}

\fig\ref{fig:rp_sd} provides an example of how RPs work and their properties. In this example, we consider two time windows with a window size of $\tau=5$, one being more fluctuating on CSI amplitude difference and the other relatively stable. The threshold value $\gamma=0.3$ is set to compute the absolute value of the difference between timestamps 1 and 3 for both windows. In \fig\ref{fig:rp_sd_a}, we consider an example of $\bar{d}^{t}_{m,n_1}=\{0.5, 0.8, 0.5, 0.2, 0.5\}$ with the absolute difference value of $x_{rp}^{3,1}=0.6$ greater than the threshold value, resulting in a white block. In contrast, in \fig\ref{fig:rp_sd_b}, we consider an example of $\bar{d}^{t}_{m,n_1}=\{0.5, 0.6, 0.5, 0.4, 0.5\}$ with the absolute difference value of $x_{rp}^{3,1}=0.2$ smaller than the threshold value, resulting in a black block. After obtaining the results for all blocks, we can see that RPs are persymmetric matrices, and their diagonals must be black because their differences are all equal to $1$. Additionally, RPs have more white blocks in more fluctuating situations and fewer white blocks in steadier ones. With this property, RPs can be adopted to efficiently distinguish between dynamic and static cases.

\subsubsection{Static Feature Extraction and Coloring CSI Ratio Image}
In this subsection, we will explain how to extract the SF and why our SF can provide better performance for NLoS case than those adopting only CSI amplitude or phase. We also design a colorization method to enhance the effectiveness of SF, whilst the channel merging can increase the data abundance with comparatively low computational complexity. The following three parts will provide detailed descriptions of each strategy.

\paragraph{\textbf{CSI Ratio and Complex Plane Image}}

\begin{figure}[t]
\centering
\includegraphics[width=3.1in]{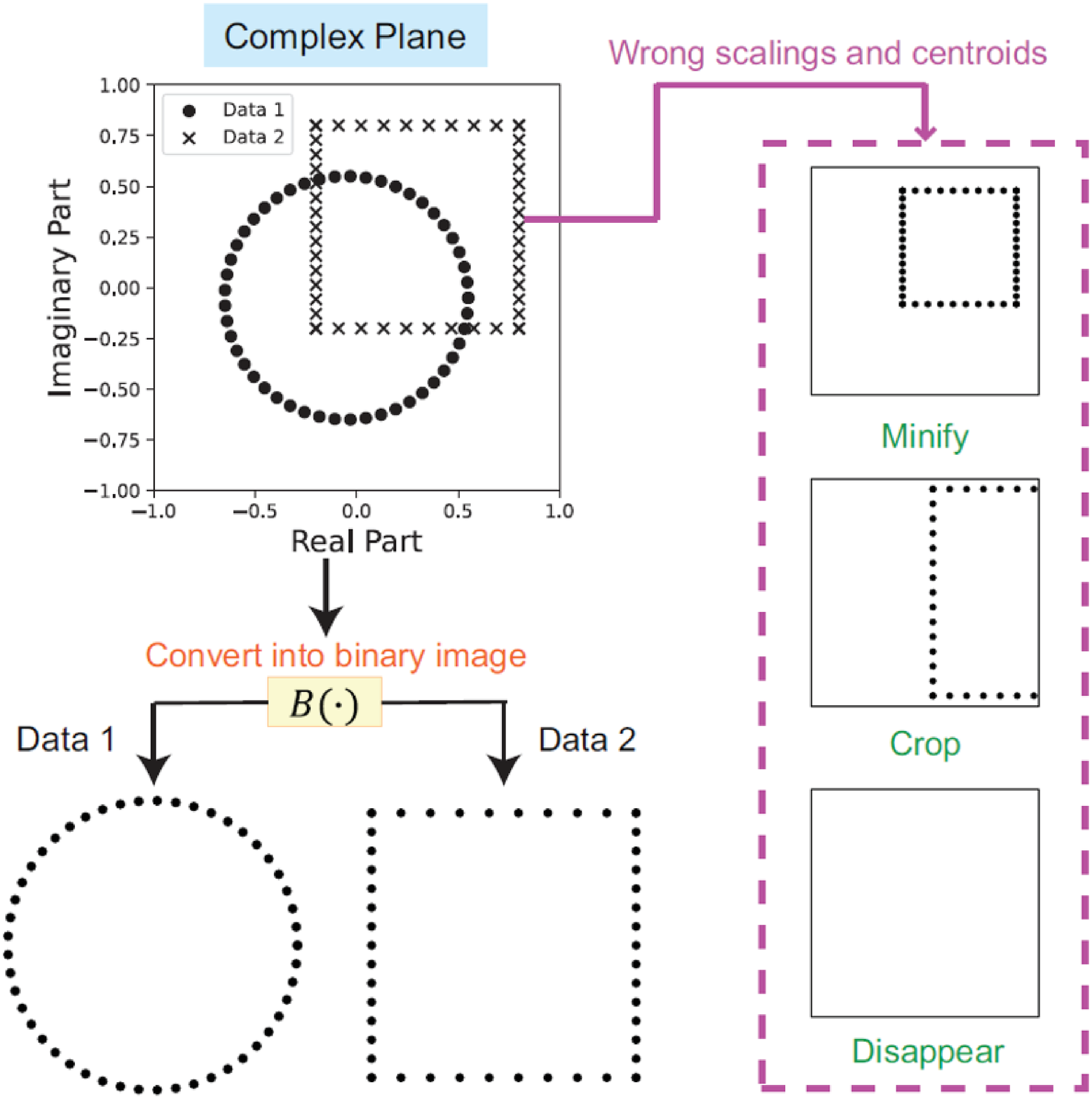}
\caption{The schematic diagram of binary CSI ratio images.}
\label{fig:ratio_sd}
\end{figure}

Many studies \cite{yuming, fangyu} on human presence detection only rely on either raw or processed CSI amplitude as the basis for classification. However, as revealed in Subsection \ref{nlos_sp}, using only CSI amplitude makes it difficult to distinguish between the signal characteristics of an empty room and an NLoS static case. The insignificant changes in CSI amplitude information can result in the false classification of these two cases. Moreover, CSI raw phase is ineffective for human sensing tasks due to the random phase offset. Nevertheless, phase information is crucial for resolving the NLoS static problem because it is more sensitive than amplitude for detecting a stationary person in the corner of the room \cite{rt_fall}. To address these issues, we employ the CSI ratio between transmission pairs in our system as the SF. The CSI ratio matrix at the last timestamp can be obtained as
\begin{equation} \label{CSIRM}
    \bm{F}^t_s = [\bm{r}^t_1, \bm{r}^t_2, \dots, \bm{r}^t_q, \dots, \bm{r}^t_Q],
\end{equation}
where $Q \le \tbinom{M \cdot N}{2}$ is the total number of the couples of transmission pairs. Note that $\tbinom{a}{b}\!=\frac{a!}{(a-b)!b!}$ is the binomial function with '$!$' as factorial operation. Each element in $\bm{F}^t_s$ in $\eqref{CSIRM}$ is expressed as
\begin{equation}
    \bm{r}^t_q = [r^t_{q,1}, r^t_{q,2}, \dots, r^t_{q,k}, \dots, r^t_{q,K}].
\label{eq:ratio_vec}
\end{equation}
Based on $\eqref{eq:csi_value}$, each element of $\bm{r}^t_q$ in $\eqref{eq:ratio_vec}$ can be formulated as
\begingroup
\allowdisplaybreaks
\begin{align}
    r^t_{q,k} &= \frac{h^t_{\alpha_1,k}}{h^t_{\alpha_2,k}} \triangleq \frac{|h^t_{\alpha_1,k}|}{|h^t_{\alpha_2,k}|}e^{j(\angle{h^t_{\alpha_1,k}}-\angle{h^t_{\alpha_2,k}})} \notag \\
    &= \frac{e^{-j\phi^t_{\alpha_1,k}}\sum_{l_{\alpha_1}=1}^{L_{\alpha_1}}A^t_{\alpha_1,l_{\alpha_1}}e^{-j2\pi\frac{d^t_{\alpha_1,l_{\alpha_1}}}{\lambda_k}}}{e^{-j\phi^t_{\alpha_2,k}}\sum_{l_{\alpha_2}=1}^{L_{\alpha_2}}A^t_{\alpha_2,l_{\alpha_2}}e^{-j2\pi\frac{d^t_{\alpha_2,l_{\alpha_2}}}{\lambda_k}}}, \notag \\
	& \overset{\underset{(a)}{}}{=} \frac{\sum_{l_{\alpha_1}=1}^{L_{\alpha_1}}A^t_{\alpha_1,l_{\alpha_1}}e^{-j2\pi\frac{d^t_{\alpha_1,l_{\alpha_1}}}{\lambda_k}}}{\sum_{l_{\alpha_2}=1}^{L_{\alpha_2}}A^t_{\alpha_2,l_{\alpha_2}}e^{-j2\pi\frac{d^t_{\alpha_2,l_{\alpha_2}}}{\lambda_k}}}.
\label{eq:ratio}
\end{align}
\endgroup
where $\alpha_1 = (m_1, n_1)$ and $\alpha_2 = (m_2, n_2)$ represent two transmission pairs, and $\alpha_1 \ne \alpha_2$. Notations of $m_1,m_2 \in [1,M]$ are the transmit antenna indices, and $n_1,n_2 \in [1,N]$ are the receiving antenna indices. We can observe that the division of CSI complex values is equivalent to dividing the amplitudes and subtracting the phases. Furthermore, the equality of (a) holds since different antennas on device share the same oscillator \cite{ranging, spotfi}. The benefit of adopting CSI ratio is that it is capable of potentially eliminating the random phase offset terms $e^{-j\phi^t_{\alpha_1,k}}$ and $e^{-j\phi^t_{\alpha_2,k}}$ since the random phase offsets for each transmission pair are considered identical, thus making phase information available.

Note that CSI is a complex value composed of real and imaginary parts. To visualize the differences between all cases, the CSI ratio vector $\bm{r}^t_q$ in $\eqref{eq:ratio_vec}$ can be plotted on a complex plane. One CSI vector has $K$ subcarrier points, and these $K$ points can form a shape on the complex plane. If all amplitude ratio and phase difference values of any two cases are different, these two CSI ratio vectors will form two distinct shapes in different areas of the plane. Therefore, the pattern of CSI ratios on the complex plane can be generated as images for classification. In our system, the SF image generating function in \fig\ref{fig:block_diagram} can generate the binary image $\bm{X}^t_{B,q}$ for each CSI vector by using the mapping function $B(\cdot)$ as
\begin{equation}
    \bm{X}^t_{B,q} = B(\bm{r}^t_q) : \mathbb{C}^{K} \to \mathbb{N}^{1 \times w \times h},
\label{eq:binary_image}
\end{equation}
which maps from complex domain $\mathbb{C}$ to integer domain $\mathbb{N}$. Notations of $w$ and $h$ indicate the width and height of the image, respectively. $\bm{X}^t_{B,q}$ is a single channel image, and subscript $B$ denotes the binary nature of the image, indicating that each pixel of an image is either 0 or 1 converting by the function of $B(\cdot)$. For example, \fig\ref{fig:ratio_sd} is a schematic diagram for understanding our designs. Data 1 and Data 2 denote two CSI ratio vectors with the number of subcarriers $K=50$. Assume that owing to changes in the wireless channel, one CSI ratio vector forms a circle pattern, and the other forms a square pattern. We can exactly observe they are distinct by their shapes. Hence, as shown in the lower-left part of \fig\ref{fig:ratio_sd}, binary images containing the shape information can be generated to classify them. The SF image generating function intends to provide customized magnitude scalings and centroids for each image. Otherwise, the shape formed by original CSI ratio points may be minified, cropped, or even disappeared when fitting wrong scalings and centroids, such as the erroneous images at the right of \fig\ref{fig:ratio_sd}. In this context, if human behaviors affect the shapes of the CSI ratios on the complex plane, binary images can be utilized as effective classification inputs.

\paragraph{\textbf{Colorization}}

\begin{figure}[t]
\centering
\includegraphics[width=3.1in]{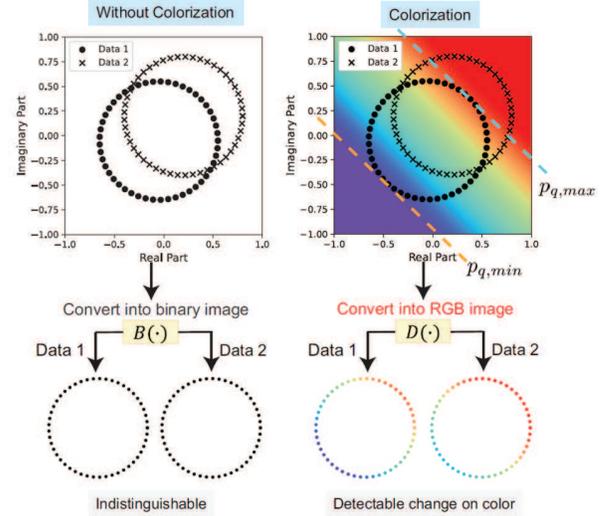}
\caption{The schematic diagram of colorization method.}
\label{fig:color_sd}
\end{figure}

After generating binary images, different cases can be distinguished if their shapes are different. However, sometimes the patterns can be very similar, and the only difference is their positions on the complex plane. As shown in the left part of \fig\ref{fig:color_sd}, if Data 2 has a slightly stronger reflection path than Data 1, all CSI ratio amplitudes of Data 2 will increase by the same value, with their phases unchanged. This will cause the CSI ratio of Data 2 to have the same shape as Data 1 but with a position offset to the upper-right corner. Notice that this situation usually happens between the empty room and static NLoS case, and the other two cases have larger fluctuations that directly change their shapes. As shown in the lower-left part of \fig\ref{fig:color_sd}, those two binary images representing two cases will not be able to be differentiated since the shapes are almost identical. The reason is that binary images only preserve shape information, and location information will be lost due to the specialized magnitude scalings and centroids when generating graphs. To address this issue, we propose a colorization algorithm that assigns colors to binary images, providing location information and enhancing the distinction between the empty room and NLoS stationary cases.

Firstly, we extract the real and imaginary parts of our location information since they represent the X-Y coordinates in a complex-domain plane. The position for each subcarrier $k$ in the CSI vector is defined as
\begin{equation}
    p^t_{q,k} = \mathfrak{R} \{r^t_{q,k}\} + \mathfrak{I} \{r^t_{q,k}\},
\end{equation}
where $\mathfrak{R}\{r^t_{q,k}\}$ and $\mathfrak{I}\{r^t_{q,k}\}$ indicate the real and imaginary parts, respectively. To ensure that the colorization scheme is effective, the range of position values needs to be determined carefully. If the range is too wide or narrow, all points will appear to be with identical color which are difficulty to be distinguished. Thus, we use the empty room case to determine the range of position values, and we calculate the time series of the average CSI ratio in case 1 as a reference. The position value for each subcarrier in the CSI vector is obtained by adding the real and imaginary parts, which can be expressed as
\begin{equation}
    \bar{\bm{r}}_q(e) = \left[\bar{r}_{q,1}(e), \bar{r}_{q,2}(e), \dots, \bar{r}_{q,k}(e), \dots, \bar{r}_{q,K}(e)\right],
\label{eq:r_bar}
\end{equation}
where
\begin{equation}
    \bar{r}_{q,k}(e) = \frac{1}{\tau_c}\sum_{\delta=0}^{\tau_c-1}r^{\delta}_{q,k}(e),
\end{equation}
where $\tau_c$ is the window size used for colorization, and $\delta$ represents the data sample index. The purpose of computing the average in this case is to obtain the general trend of all subcarrier values during the considered time period. Once we have computed the average, we search for the maximum and minimum position values from the $\bar{\bm{r}}_q(e)$ in $\eqref{eq:r_bar}$ by
\begin{equation} \label{pp1}
    p_{q,max} = \max_{k}\left( \mathfrak{R} \{\bar{\bm{r}}_q(e)\}+ \mathfrak{I} \{\bar{\bm{r}}_q(e)\} \right),
\end{equation}
\begin{equation} \label{pp2}
    p_{q,min} = \min_{k}\left( \mathfrak{R}\{\bar{\bm{r}}_q(e)\}+ \mathfrak{I}\{\bar{\bm{r}}_q(e)\}\right),
\end{equation}
where $\max_{k}$ and $\min_{k}$ obtain the maximum and minimum values along the subcarrier index $k$, respectively. By limiting the range of position values with $p_{q,max}$ and $p_{q,min}$, most of the position values in case 1 fall within this range. This allows us to transform the position value of each CSI ratio point into a color, and the resulting color matrix for the $q$-th couple can be represented as
\begin{equation}
    \bm{C}^t_q = \left[\bm{c}^t_{q,1}, \bm{c}^t_{q,2}, \dots, \bm{c}^t_{q,k}, \dots, \bm{c}^t_{q,K}\right], 
\end{equation}
where
\begin{equation}
    \bm{c}^t_{q,k} = 
    \begin{cases}
    \bm{c}_{q,max}, & \mbox{if } p^t_{q,k} \ge p_{q,max,}\\
    f\left(p^t_{q,k}\right), & \mbox{if } p_{q,min} < p^t_{q,k} < p_{q,max},\\
    \bm{c}_{q,min}, & \mbox{if } p^t_{q,k} \le p_{q,min},
    \end{cases}
\end{equation}
with $f(\cdot)$ as the mapping function from real domain $\mathbb{R}$ to integer domain $\mathbb{N}^3$. Accordingly, the red-green-blue (RGB) image of the $q$-th CSI ratio vector can be generated as
\begin{equation}
    \bm{X}^t_{RGB,q} = D(\bm{X}^t_{B,q}, \bm{C}^t_q): (\mathbb{N}^{1\times w \times h}, \mathbb{N}^{3\times K}) \to \mathbb{N}^{3 \times w \times h},
\end{equation}
where $D(\cdot)$ is the mapping function that maps the binary image into RGB-based images, with three dimensions in $\bm{X}^t_{RGB,q}$ representing the RGB channel. The RGB images not only capture the shape characteristics but also provide location information through color, making $\bm{X}^t_{RGB,q}$ a superior representation compared to the binary image $\bm{X}^t_{B,q}$ in addressing the NLoS static problem. For example, the rainbow spectrum at the right side of \fig\ref{fig:color_sd} is employed as our color bar, where $\bm{c}_{q,max}=f(p_{q,max})$ is defined to be red and $\bm{c}_{q,min}=f(p_{q,min})$ is purple. If the position value exceeds the defined upper or lower bound, it will be colored in red or purple, respectively. Since the range of the color bar is based on the empty room case, the images in case 1, such as Data 1 in \fig\ref{fig:color_sd}, will have a more uniform distribution of colors from red to purple. In contrast, Data 2 is slightly shifted to the upper-right area with some points exceeding the boundary. It can be observed that the resulting image of Data 2 possess more points with red color. Therefore, even though their shapes are identical, our colorization method can amplify their visual difference to become distinguishable.

\paragraph{\textbf{Channel Merging}}

After obtaining the RGB images of all $Q$ couples, the classification input consists of the combination of these $Q$ images. To reduce computational complexity, we convert the images into low-resolution images $\bm{X}^t_{G,q}$ utilizing a mapping function $G(\cdot)$ as
\begin{equation}
    \bm{X}^t_{G,q} = G(\bm{X}^t_{RGB,q}): \mathbb{N}^{3\times w \times h} \to \mathbb{N}^{1 \times w \times h}.
\end{equation}
The reason for transforming RGB to grayscale based images is due to the benefits of comparatively smaller data size and retainability of colorization. Note that this method of grey images can still achieve the desired effect of colorization. Notice that it is infeasible for the original binary images to be directly converted to grey images because the upper and lower bounds of grayscale are fixed to be white and black, respectively. The classifying cases can become distinguishable if most data occurs in the area of white color, for example. The colorization scheme, on the other hand, provides feasible upper and lower bounds as in $\eqref{pp1}$ and $\eqref{pp2}$, respectively, which can effectively adjust the considered data ranges. Thus, we generate RGB images during colorization and then convert them to greyscale via channel merging. Finally, the grey images of all $Q$ couples can be merged as follows:
\begin{equation}
    \bm{X}^t_G = \left[\bm{X}^t_{G,1},\bm{X}^t_{G,2},\dots,\bm{X}^t_{G,q},\dots,\bm{X}^t_{G,Q}\right].
\end{equation}
Note that combining the CSI ratios of different couples in $\bm{X}^t_G$ increases the richness of data. Our classification model takes $\bm{X}^t_{ratio} = \frac{1}{s}\bm{X}^t_G$ as input, where $s$ is the maximum grayscale value for each pixel, and $t$ represents the final timestamp in the time window. We refer to $\bm{X}^t_{ratio}$ as the merged coloring CSI ratio image, which includes the effect of colorization as well as the information between different couples for classifying static cases.

\begin{figure}
\centering
\includegraphics[width=3.5in]{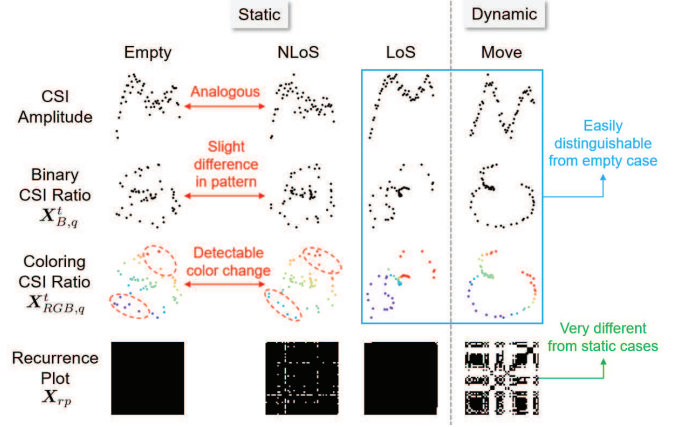}
\caption{Exemplified input images for human presence detection.}
\label{fig:input_image}
\end{figure}

In Fig. \ref{fig:input_image}, we present an example for different input images for the four cases, including the CSI amplitude image, binary CSI ratio image $\bm{X}^t_{B,q}$, coloring CSI ratio image $\bm{X}^t_{RGB,q}$, and RP $\bm{X}_{rp}$. The CSI amplitude image is generated from a transmission pair, where the horizontal axis in image represents 56 subcarriers, and the vertical axis represents their corresponding amplitude values. $\bm{X}^t_{B,q}$, $\bm{X}^t_{RGB,q}$, and $\bm{X}_{rp}$ are generated by FEIG, where the CSI ratio only contains one couple. It is worth mentioning that RP is a graph that retains information about a time series, while the other three are pictures that only contain information in one timeslot. We focus first on the empty room and NLoS static case. We can observe that the shapes in the CSI amplitude images are similar between these two cases. The reason is that multipath signals reflected by the stationary person in the corner have a long propagation path and will be absorbed by the human body, which causes subtle changes to CSI dominated by the LoS signal. Therefore, using only the amplitude is insufficient to distinguish between these two cases. Next, we examine the binary CSI ratio images of the empty and NLoS cases. They have only a slight difference in pattern, indicating that the binary image only containing the shape information cannot distinguish the NLoS situation from the empty case. Furthermore, because the colorization algorithm assigns the positional information of the complex plane to each point as a color, the coloring CSI ratio images have detectable color changes. The coloring image in the empty case possesses similar area size for each color, while the NLoS static case has more red color points without any purple points. Thus, coloring CSI ratio images for classification can effectively resolve the NLoS static problem.

On the other hand, since either standing at LoS or walking persons will have a more significant influence on the CSI signal, the CSI amplitude and ratio in these two cases will be different from that of the empty case. Hence, it is easy to distinguish them from the absence of a human. Finally, we can infer from the RPs that the walking case is highly distinguishable from the other three static cases. The walking situation will cause the amplitude difference of a time window to fluctuate, leading to more white areas in the RP. The CSI amplitude difference of the other three static cases is relatively stable, making their RPs almost black. Therefore, RPs can be employed to distinguish between dynamic and static cases. After merging all couples of CSI ratios through channel merging, we input the RPs and merged coloring CSI ratio images in Fig. \ref{fig:input_image} into our three-stage contrastive model to predict the final human presence detection result, as elaborated in the following subsection.

\subsection{Three-Stage Supervised Contrastive Learning}

In this subsection, we will introduce our three-stage supervised contrastive learning as shown in \fig\ref{fig:block_diagram}, where inputs are $\bm{X}_{rp}$ and $\bm{X}^t_{ratio}$. Our supervised contrastive learning framework is inspired by SupCon \cite{supcon} and aims to learn representations that can discriminate between different classes in three stages, which are respectively elaborated as follows.

%
%

\subsubsection{\textbf{Stage 1 - RP-based Supervised Contrastive Learning}}

We use supervised contrastive learning to train the RP encoder. The goal of contrastive learning is to bring similar samples closer and push dissimilar samples apart in the embedding space, which enables us to cluster data with similar features and separate groups with different characteristics. The term supervised means we incorporate label information into the contrastive learning process, so that positive samples with the same label as the anchor are brought closer, while negative samples with different labels are pushed away. This allows us to group representations with the same label and separate those with different labels, which can be useful for classification using a simpler model.

\begin{figure*}
  \centering
  \begin{subfigure}[b]{0.27\textwidth}
    \includegraphics[width=\textwidth]{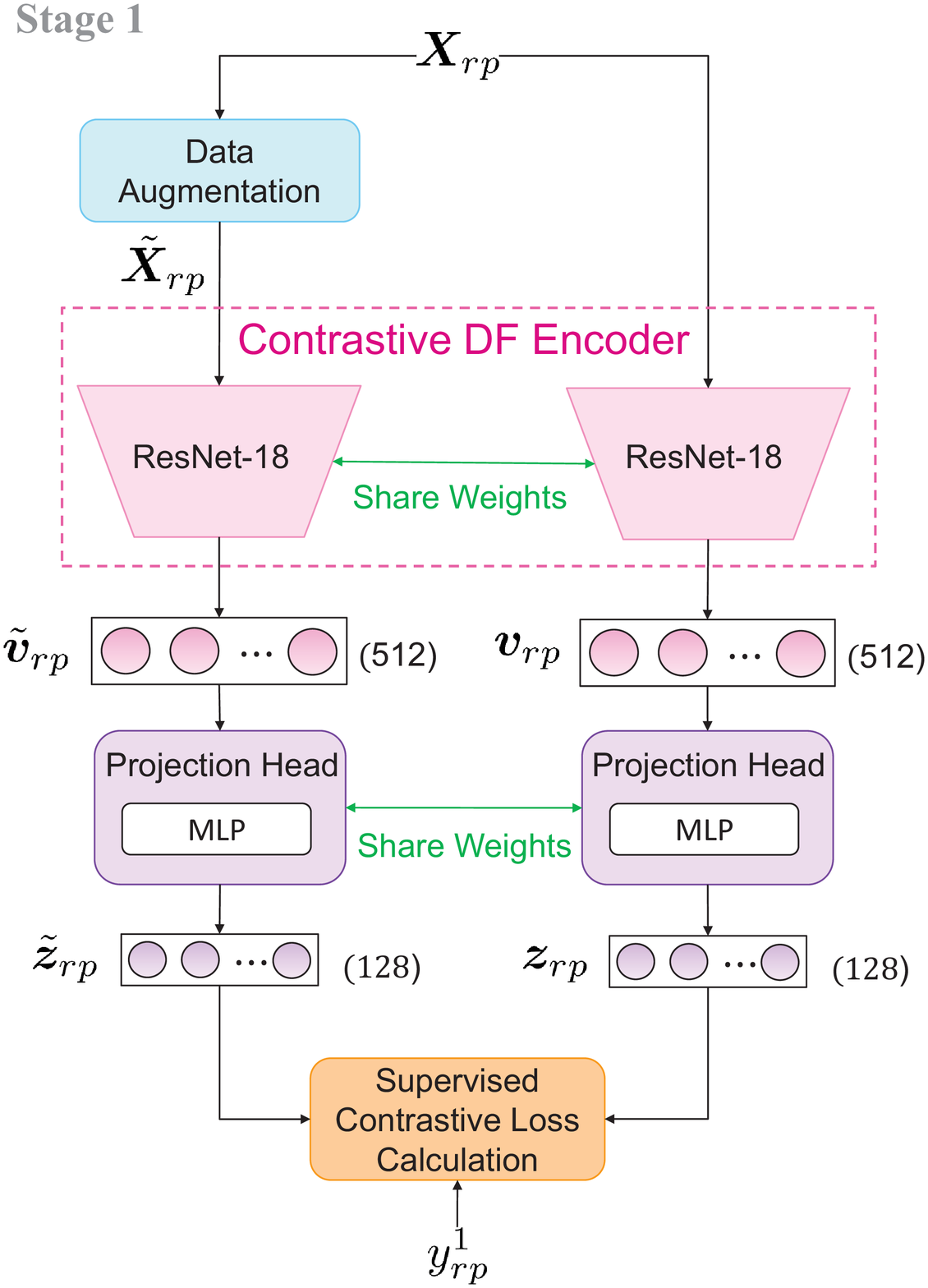}
    \caption{\footnotesize}
    \label{fig:s1_model}
  \end{subfigure}
  \begin{subfigure}[b]{0.4\textwidth}
    \includegraphics[width=\textwidth]{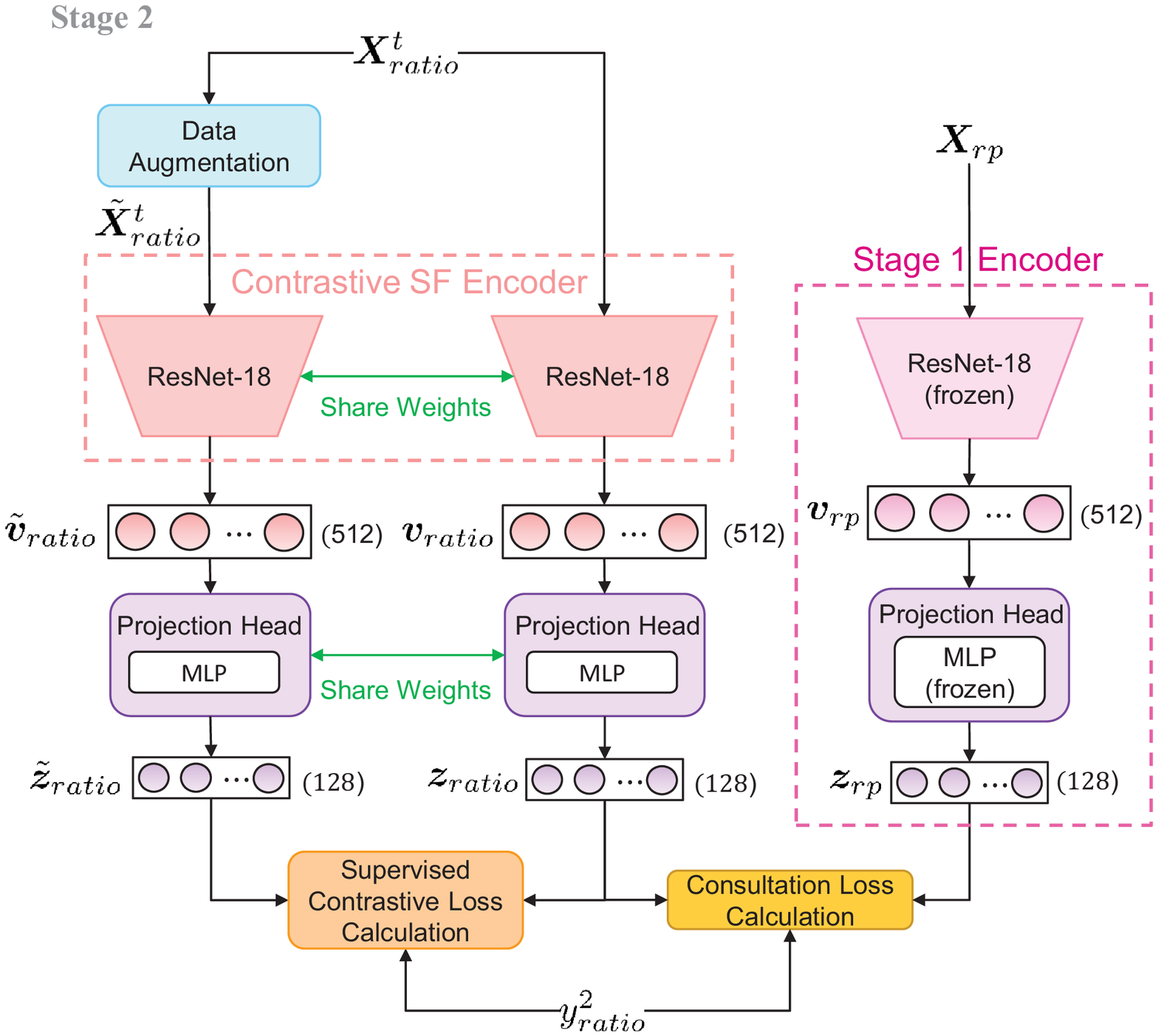}
    \caption{\footnotesize}
    \label{fig:s2_model}
  \end{subfigure}
  \begin{subfigure}[b]{0.24\textwidth}
    \includegraphics[width=\textwidth]{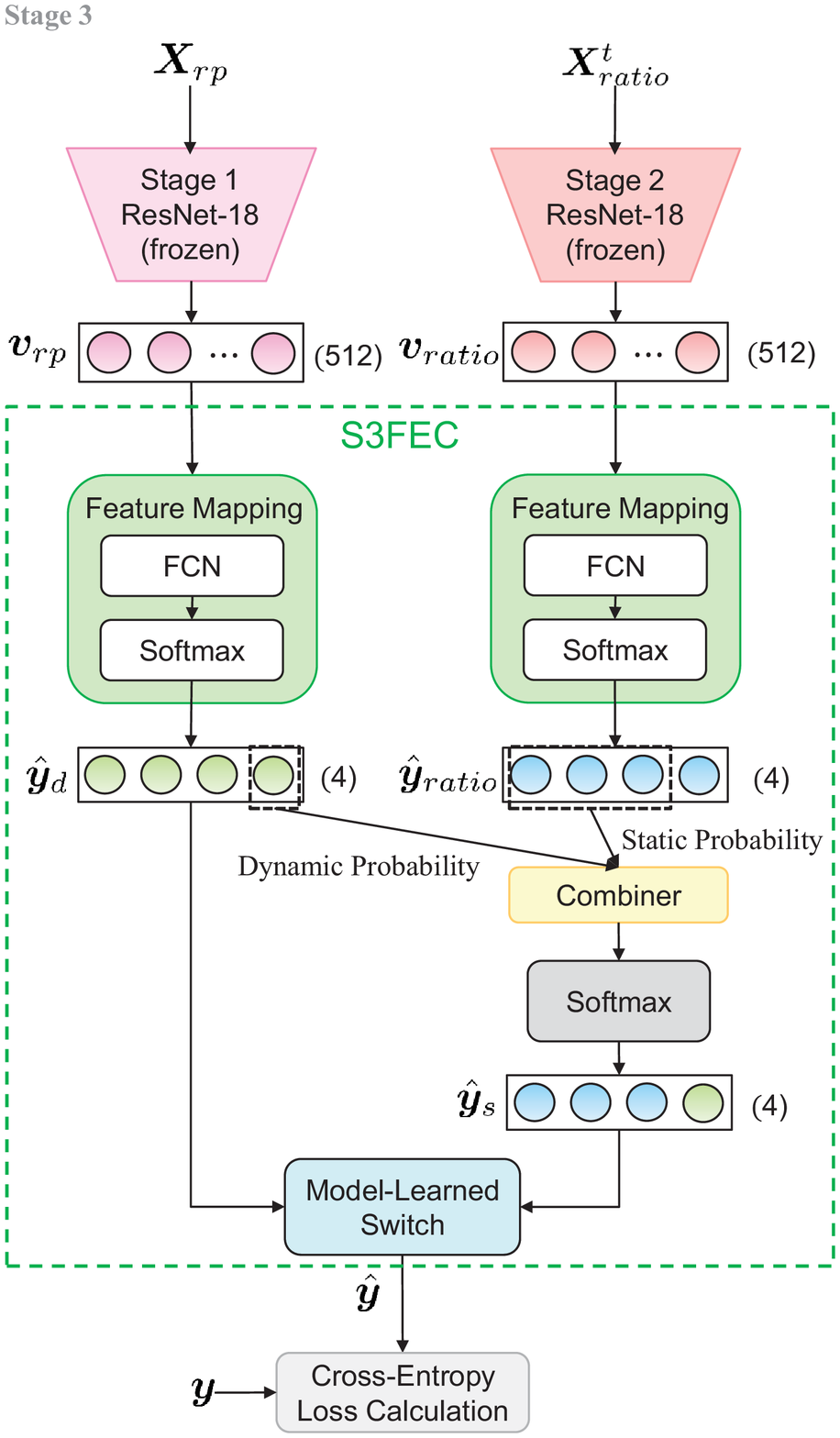}
    \caption{\footnotesize}
    \label{fig:s3_model}
  \end{subfigure}
\caption{The learning modules of (a) stage 1, (b) stage 2 and (c) stage 3.}
\end{figure*}

\par

\fig\ref{fig:s1_model} depicts a flow diagram of our neural network in stage 1. First, data augmentation is executed for the original RP to generate the augmented sample, which can be presented as
\begin{equation}
    \tilde{\bm{X}}_{rp} = Aug(\bm{X}_{rp}): \mathbb{N}^{1 \times w \times h} \to \mathbb{N}^{1 \times w \times h},
\end{equation}
where $Aug(\cdot)$ is responsible for data augmentation, which provides different views of the original image and improves the model generalization. Since RPs are binary images, our data augmentation operations only involve random resize cropping with a scale of $(\epsilon_1, \epsilon_2)$ and random horizontal flipping with a probability of $\varepsilon$. After data augmentation, both the raw and augmented images can be passed through the same contrastive DF encoder. The representation of original sample can be formulated as
\begin{equation}
    \bm{v}_{rp} = ResNet^1(\bm{X}_{rp}): \mathbb{N}^{1 \times w \times h} \to \mathbb{R}^{512},
\end{equation}
where $ResNet^1$ is ResNet-18 \cite{resnet} performed as a contrastive DF encoder, and superscript 1 means stage 1. The representation vector of an augmented image can be expressed as $\tilde{\bm{v}}_{rp} = ResNet^1(\tilde{\bm{X}}_{rp})$. Notation $\bm{v}_{rp}$ and $\tilde{\bm{v}}_{rp}$ in our work are vectors with 512 dimensions representing the feature of their images. Afterwards, the projection head will project the representation to a vector with lower dimension expressed as
\begin{equation}
    \bm{z}_{rp} = Norm \Big( MLP^1(\bm{v}_{rp}) \Big) = Norm \Big( \bm{W}_2\, \sigma(\bm{W}_1 \bm{v}_{rp}) \Big),
\end{equation}
where $Norm(\cdot)$ is normalization function, whilst $MLP^1(\cdot)$ means the multilayer perceptron (MLP) in stage 1. $\bm{W}_1$ and $\bm{W}_2$ are weight matrices with respective dimensions of $512 \times 512$ and of $128 \times 512$. Accordingly, the output performs a vector with a dimension of $128$. $\sigma(x) = \max(0,x)$ is defined as the activation function of rectified linear unit (ReLU). Moreover, the projection output of $\tilde{\bm{v}}_{rp}$ can be presented as $\tilde{\bm{z}}_{rp} = Norm ( MLP^1(\tilde{\bm{v}}_{rp}) )$. 
Notice that performing projection head is beneficial for contrastive loss as stated in \cite{simclr}. 

Eventually, the supervised contrastive loss in stage 1 can be calculated based on $\bm{z}_{rp}$ and $\tilde{\bm{z}}_{rp}$. We define $B$ as the data batch size. Therefore, a batched data with a size of $2B$ can be denoted as $\{\bm{z}^1_{rp,b},y^1_{rp,b}\}_{b=1,\dots,2B}$, where $\bm{z}^1_{rp,b}$ is composed of the original and augmented samples. The notation $y^1_{rp,b}$ denotes the ground-truth label, where the label information of augmented data is the same as that of the original data with superscript 1 signifies stage 1. Subsequently, the supervised contrastive loss can be obtained as
\begin{equation}
    L_{sc} = \sum_{i\in I} \frac{-1}{\mathcal{C}(P(i))} \sum_{p \in P(i)} \log\frac{\exp(\bm{z}^1_{rp,i}\cdot \bm{z}^1_{rp,p}/\zeta)}{\sum\limits{a \in A(i)} \exp(\bm{z}^1_{rp,i}\cdot \bm{z}^1_{rp,a}/\zeta)},
\label{eq:sc_loss}
\end{equation}
where $I=\{1,2,\dots,2B\}$ indicates an integer set with a length of $2B$. In $\eqref{eq:sc_loss}$, $P(i) = \{p \in A(i) |\forall y^1_{rp,p} = y^1_{rp,i}\}$ includes the indices of all positives, $A(i) = I \backslash i$ denotes the index set of the remaining samples except the $i$-th data sample. The function of $\mathcal{C}(P(i))$ represents the cardinality of $P(i)$, whereas $\zeta$ is the scalar temperature parameter. As observed in $\eqref{eq:sc_loss}$, the numerator and denominator represent the inner product of $i$-th anchor sample and positive ones as well as the inner product of anchor and the rest of data, respectively. The inner product measures the similarity between two vectors, which is also indicative of their distance in the projection space. As the loss decreases, the inner product in the numerator will approach the maximum value of 1, and the value of the denominator will decrease. This indicates that the loss will make the data with the same label closer, while the data with different labels will become more distinct. After the training process of stage 1, the contrastive DF encoder can learn to output discriminative representations. Note that the well-trained $\bm{v}_{rp}$ can be input to the model in stage 3 to classify the four cases of human presence detection.

\subsubsection{\textbf{Stage 2 - CSI-based Supervised Contrastive Learning and Consultation Loss}}


In stage 2, we focus on supervised contrastive learning for merged coloring CSI ratio images $\bm{X}^t_{ratio}$, which is  demonstrated in \fig\ref{fig:s2_model}. Notice that the operation on the left is the same as stage 1. First, the augmented sample of $\bm{X}^t_{ratio}$ can be obtained by $\tilde{\bm{X}}^t_{ratio} = Aug(\bm{X}^t_{ratio}): \mathbb{R}^{Q \times w \times h} \to \mathbb{R}^{Q \times w \times h}$, and the data augmentation function also adopts random resize cropping and random horizontal flipping here. Next, the representations of original and augmented samples can be transformed as $\bm{v}_{ratio} = ResNet^2(\bm{X}^t_{ratio}): \mathbb{R}^{Q \times w \times h} \to \mathbb{R}^{512}$ and $\tilde{\bm{v}}_{ratio} = ResNet^2(\tilde{\bm{X}}^t_{ratio})$, respectively, where $ResNet^2$ is also ResNet-18 with input channel equal to $Q$ as the contrastive SF encoder, and superscript 2 indicates stage 2. Afterwards, their projection outputs are obtained as $\bm{z}_{ratio} = Norm (MLP^2(\bm{v}_{ratio}) )$ and $\tilde{\bm{z}}_{ratio} = Norm (MLP^2(\tilde{\bm{v}}_{ratio}))$, where the $MLP^2(\cdot)$ has the identical structure as $MLP^1(\cdot)$. Before computing the supervised contrastive loss, we additionally define the batched data in stage 2 as $\{ \bm{z}^2_{ratio,b}, y^2_{ratio,b}\}_{b=1,\dots,2B}$. Again, $\bm{z}^2_{ratio,b}$ contains the original and augmented data samples. Finally, the supervised contrastive loss can be computed similar to $\eqref{eq:sc_loss}$ based on $\bm{z}^2_{ratio,b}$ and $y^2_{ratio,b}$.

However, due to the challenge induced by the NLoS static problem, we need to introduce an additional design to further separate the representations of the empty and NLoS cases in the embedding space. After training the contrastive DF encoder in stage 1, the representations of the dynamic and static cases should be separated by distance. Since we intend to use $\bm{X}^t_{ratio}$ to classify the three static situations, expanding the distances between the representation vectors of these three cases is expected to be achieved. Therefore, the consultation loss is designed to aid the supervised contrastive loss during training, which is given by
\begin{align}
    L_{cs} &= \left| \frac{1}{\mathcal{C}(S_s)}\sum_{(i,j)\in S_s} d(\bm{z}^2_{ratio,i}, \bm{z}^2_{ratio,j})  \right. \notag\\
    & \left. \qquad\qquad- \frac{1}{\mathcal{C}(S_{ds})}\sum_{(i,j)\in S_{ds}} d(\bm{z}^c_{rp,i}, \bm{z}^c_{rp,j}) \right|,
\label{eq:cs_loss}
\end{align}
where $S_s \!=\! \{i,j \in [1, B]| \forall i\!\ne\! j\}$ satisfying cases and labels of $\left\lbrace\left(y^2_{ratio,i}, y^2_{ratio,j}\right) =(1,2)\right\rbrace \cup \left\lbrace\left(y^2_{ratio,i}, y^2_{ratio,j}\right)=(1,3) \right\rbrace \cup \left\lbrace\left(y^2_{ratio,i}, y^2_{ratio,j}\right)=(2,3) \right\rbrace$ represents the unified set for the pairs of three static cases within a contrastive data batch. Note that $\cup$ stands for union operation among sets, subscript $s$ means stationary, and the values of labels correspond to the four cases mentioned in Subsection \ref{sys_arc}. $S_{ds}(i,j)= \{i,j \in [1, B]|\forall i\ne j\}$ is the index set that meets $\left\lbrace\left(y^2_{ratio,i}, y^2_{ratio,j}\right) =(1,4)\right\rbrace \cup \left\lbrace\left(y^2_{ratio,i}, y^2_{ratio,j}\right)=(2,4) \right\rbrace \cup \left\lbrace\left(y^2_{ratio,i}, y^2_{ratio,j}\right)=(3,4) \right\rbrace$ comparing static with dynamic cases. The subscript $ds$ denotes joint dynamic and static cases. Moreover, $d(\cdot)$ is the function calculating the Euclidean distance between two vectors. The first term of $\eqref{eq:cs_loss}$ represents the average distance between the projection vectors $\bm{z}_{ratio}$ among three static situations, while the second term means that of $\bm{z}_{rp}$ between dynamic and static cases with the superscript $c$ denoting consultation. Note that the \textit{consultation} loss is named for the reason that it involves consulting the stage 1 encoder when training the contrastive SF encoder. In $\eqref{eq:cs_loss}$, we can observe that as the loss decreases both terms become closer. This indicates that the distance between the projection vectors of $\bm{X}^t_{ratio}$ for the three static cases is escalated, allowing us to obtain representative CSI information to classify these three cases.

As shown on the right side of \fig\ref{fig:s2_model}, both ResNet-18 and the projection head are essential for stage 2 training, and the stage 1 encoder is necessary to provide the well-trained model between dynamic and static RPs. A data batch from the stage 1 encoder can be defined as $\{\bm{z}^c_{rp,b},y^2_{ratio,b}\}_{b=1,\dots,B}$. Note that this dataset does not contain the augmented data because we only require the reference distance. The label $y^2_{ratio,b}$ corresponds to each merged coloring CSI ratio image in the dataset. Therefore, we can calculate the consultation loss from $\bm{z}_{rp}$ and $\bm{z}_{ratio}$, which sustains the asymptotic impact on representation since projection head is a dimension reduction function. Finally, leveraging supervised contrastive loss in $\eqref{eq:sc_loss}$ and consultation loss in $\eqref{eq:cs_loss}$, we can derive the total loss of stage 2 as
\begin{equation}
	L_{s2} = L_{sc} + \lambda \cdot L_{cs},
\end{equation}
where the constant $\lambda$ controls the significance of the consultation loss. After the completion of stage 2 training, we can obtain improved representations of merged coloring CSI ratio images, which can be employed to address the NLoS static problem in stage 3 by combining the supervised contrastive loss and the consultation loss.

\subsubsection{\textbf{Stage 3 - Self-Switched Static Feature Enhanced Classifier (S3FEC)}}
\label{stage3}

In the final stage 3, our goal is to classify the representations learned from stages 1 and 2. Typically, after representation learning, a single representative vector can be efficiently classified by a simpler model such as fully-connected neural (FCN) networks and MLP \cite{simclr, STF-CSL}. However, our system involves two representations, $\bm{X}_{rp}$ and $\bm{X}^t_{ratio}$, which requires a multi-modal representation learning mechanism. While the joint representation (JR) method that concatenates the representation vectors between different views has achieved remarkable success, it cannot resolve the differences between our two representations. RPs are designed to be advantageous for distinguishing between dynamic and static cases, but they are not well-suited for differentiating among the three static cases. On the other hand, merged coloring CSI ratio images are generated to classify between static situations, but they result in misclassification between moving and standing cases due to their limited timestamp information. Therefore, directly concatenating these two representations for classification will cause them to restrict each other deteriorating overall performance.

To address this issue, we have designed the S3FEC, which leverages the joint model of RPs for the moving case and coloring CSI ratios for static ones. \fig\ref{fig:s3_model} illustrates the network flow diagram of stage 3. We employ the two trained ResNet-18 encoders from stages 1 and 2 to generate representations. The feature mapping converts the representation vector into classification probability through one FCN and softmax activation function. Accordingly, the classification probability of $\bm{X}_{rp}$ can be expressed as
\begin{equation}
    \hat{\bm{y}}_{d} = [\hat{y}_{d,1}, \hat{y}_{d,2}, \hat{y}_{d,3}, \hat{y}_{d,4}]^T = \rho(\bm{v}_{rp} \bm{W}_{rp}+\bm{b}_{rp}),
\end{equation}
where subscript $d$ indicates the dynamic, and the four elements denote the predicted probabilities of four cases from RP. $\rho(\cdot)$ is the softmax activation function, and $\bm{W}_{rp}$ and $\bm{b}_{rp}$ are the weight and bias, respectively, in FCN for $\bm{v}_{rp}$. Similarly, the classification probability of $\bm{X}^t_{ratio}$ can be formulated as
\begin{align}
    \hat{\bm{y}}_{ratio} &= [\hat{y}_{ratio,1}, \hat{y}_{ratio,2}, \hat{y}_{ratio,3}, \hat{y}_{ratio,4}]^T \notag\\
    &= \rho(\bm{v}_{ratio} \bm{W}_{ratio}+\bm{b}_{ratio}).
\label{eq:p_ratio}
\end{align}
The weight and bias of FCN for $\bm{v}_{ratio}$ in $\eqref{eq:p_ratio}$ are denoted as $\bm{W}_{ratio}$ and $\bm{b}_{ratio}$, respectively. The outputs of $\bm{v}_{rp}$ and $\bm{v}_{ratio}$ provide two kinds of classification probabilities for the final classification. 

However, the key challenge is how to choose the appropriate probability for classification. The S3FEC introduces a model-learned switch that automatically selects the final classification probability. Since the characteristics of RPs vary significantly in dynamic and stationary scenarios, RPs are more accurate in distinguishing these two categories. Ideally, $\hat{y}_{d,4}$ should be the largest element in the dynamic case, whereas is should be the smallest element in the static situations. Thus, if $\hat{y}_{d,4}$ is the maximum probability, the switch selects $\hat{\bm{y}}_{d}$ as the final classification probability. Otherwise, if $\hat{y}_{d,4}$ is not the maximum value, RPs consider this situation a static case. The switch selects the probability generated by $\bm{X}^t_{ratio}$ for classification. However, directly taking $\hat{\bm{y}}_{ratio}$ as the final probability may cause misclassification between moving and standing cases. To address this issue, static feature enhancement is employed. The probability of case 4 predicted by $\bm{v}_{ratio}$ is replaced by the 4-th element of $\bm{v}_{rp}$ because RPs can detect no human walking in the room with probability of case 4 relatively low. By combining the probabilities from dynamic/static cases of $\hat{\bm{y}}_{d}$ and of $\hat{\bm{y}}_{ratio}$, the final classification probability vector in stage 3 can be represented by
\begin{equation}
    \hat{\bm{y}}'_{ratio} = \left[\hat{y}_{ratio,1}, \hat{y}_{ratio,2}, \hat{y}_{ratio,3}, \hat{y}_{d,4}\right]^T.
\end{equation}
The softmax activation function is employed for re-scaling $\hat{\bm{y}}'_{ratio}$ as a probability vector, which is given by $\hat{\bm{y}}_{s} = \rho(\hat{\bm{y}}'_{ratio})$. Therefore, the final classification probability can be formulated as
\begin{equation}
    \hat{\bm{y}} = \left[\hat{y}_{1}, \hat{y}_{2}, \hat{y}_{3}, \hat{y}_{4}\right]^T = \omega \hat{\bm{y}}_{d} + (1-\omega)\hat{\bm{y}}_{s},
\end{equation}
where $\omega$ means the model-learned switch value. The model-learned switch determines whether the dynamic probability from $\hat{\bm{y}}_{d}$ is the maximum value. The switch $\omega \rightarrow 1$ with $\hat{\bm{y}}_{d}$ is considered as the final classification probability. Otherwise, $\hat{\bm{y}}_{s}$ becomes the final classification probability when $\omega \rightarrow 0$. Accordingly, the final prediction is conduced with the employment of cross-entropy loss as
\begin{equation}
    L_{ce} = -\sum^B_{i=1}\sum^3_{j=0}y_{j}(i)\log\hat{y}_{j}(i).
\end{equation}
To elaborate a little further, the block diagram shown in Fig. \ref{fig:block_diagram} includes the well-trained contrastive DF and SF encoders, as well as the online parameters of S3FEC in stage 3 obtained after offline training. In the online phase, after collecting the CSI data, real-time $\bm{X}_{rp}$ and $\bm{X}^t_{ratio}$ can be generated by FEIG, as described in Subsection \ref{FEIG}. The data can then be fed into the online three-stage contrastive learning model, whereas the final output $y_{pre}$ can be predicted by our system for human presence detection.

\begin{table}[t]
\scriptsize
\centering
\caption{Training/Testing Data Size}
\begin{tabular}{|c||c|c|c|c|}
\hline
\multirow{2}{*}{}& \multicolumn{2}{c|}{Scenario 1} & \multicolumn{2}{c|}{Scenario 2} \\ \cline{2-5}
& Training & Testing & Training & Testing \\ \hline
Case 1 & 2000 & 2000 & 1000 & 1000\\ \hline
Case 2 - upper left corner & 1000 & 500 & 500 & 500 \\ \hline
Case 2 - lower right corner & 1000 & 500 & 500 & 500 \\ \hline
Case 3 & 1000 & 500 & 0 & 0 \\ \hline
Case 4 & 2000 & 1000 & 1000 & 1000\\ \hline
\end{tabular}
\label{data_sz}
\end{table}

\begin{table}[t]
\centering
\scriptsize
\caption{System Parameters}
\begin{tabular}{|l|l|}
\hline
Parameters & Value \\ \hline \hline
Carrier frequency & $2.447$ GHz \\ \hline
Channel bandwidth & $20$ MHz \\ \hline
Number of transmitter/receiver antennas $\{M, N\}$ & $\{2,2\}$ \\ \hline
Number of subcarriers $K$ & 56 \\ \hline
Sampling rate of data collection& 10 Hz \\ \hline
Window size of RPs $\tau$ & 50 \\ \hline
Window size for the threshold of RPs $\tau_{\gamma}$ & $\{2000, 1000\}$ \\ \hline
Threshold of RPs $\gamma$ & $\{10, 33\}$ \\ \hline
Number of transmission pairs $Q$ & $\{3, 2\}$ \\ \hline
Size of image $w\times h$ & $32\times 32$ \\ \hline
Window size for colorization $\tau_c$ & $\{2000, 1000\}$ \\ \hline
Maximum value of each pixel $s$ & 255 \\ \hline
Scale of random resize cropping $(\epsilon_1, \epsilon_2)$ & (0.2, 1) \\ \hline
Probability of random horizontal flipping $\varepsilon$ & 0.5 \\ \hline
Batch size $B$ & 128 \\ \hline
Epoch for stages 1 and 2 & 30 \\ \hline
Epoch for stage 3 & 10 \\ \hline
Scalar temperature parameter $\zeta$ & 0.07 \\ \hline
Weight for consultation loss $\lambda$ & 0.5 \\ \hline
 
\end{tabular}
\label{parameter}
\end{table}

\begin{figure}[t]
  \centering
  \begin{subfigure}[b]{0.22\textwidth}
    \includegraphics[width=\textwidth]{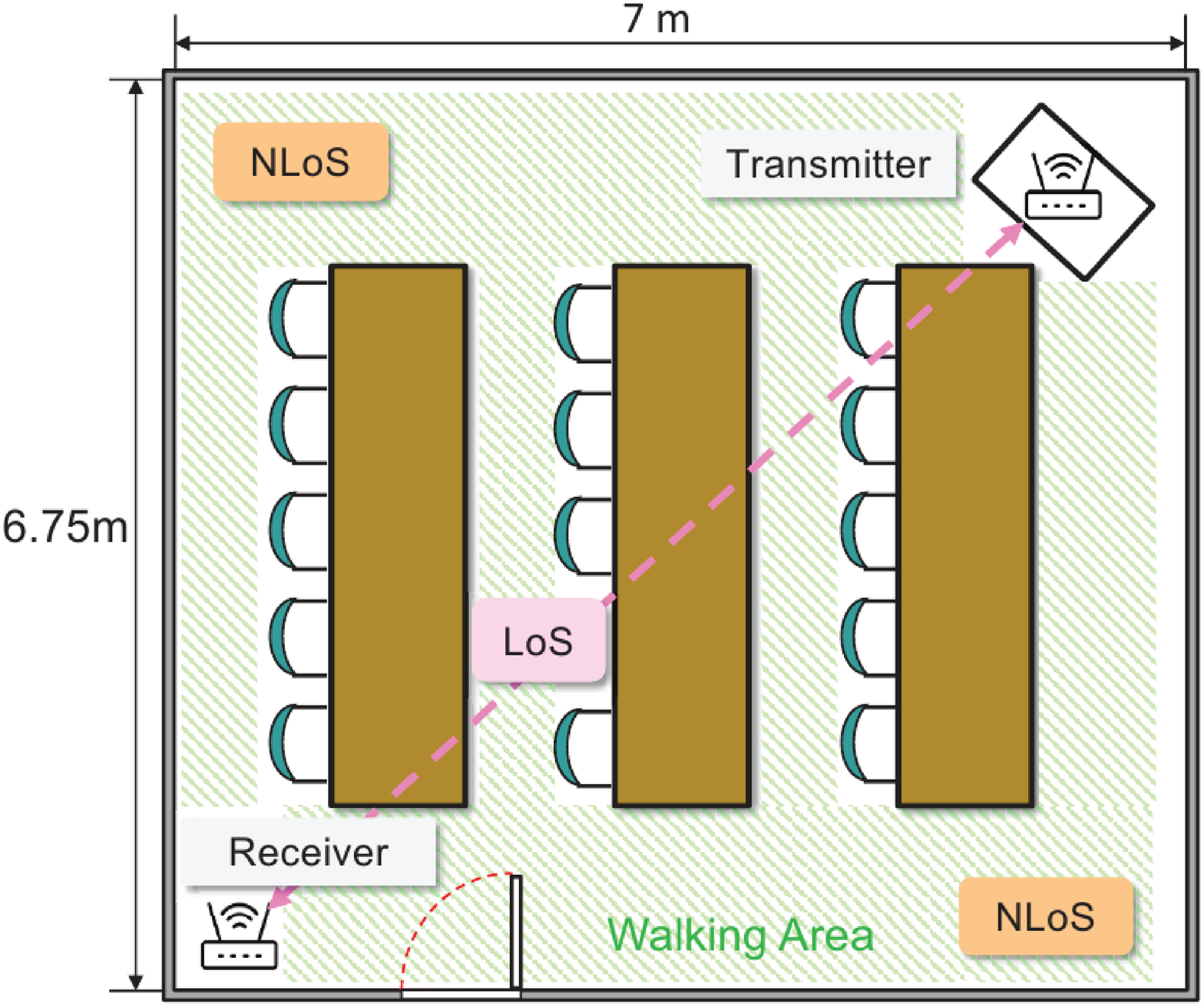}
    \caption{\footnotesize}
    \label{fig:801_layout}
  \end{subfigure}
  \begin{subfigure}[b]{0.23\textwidth}
    \includegraphics[width=\textwidth]{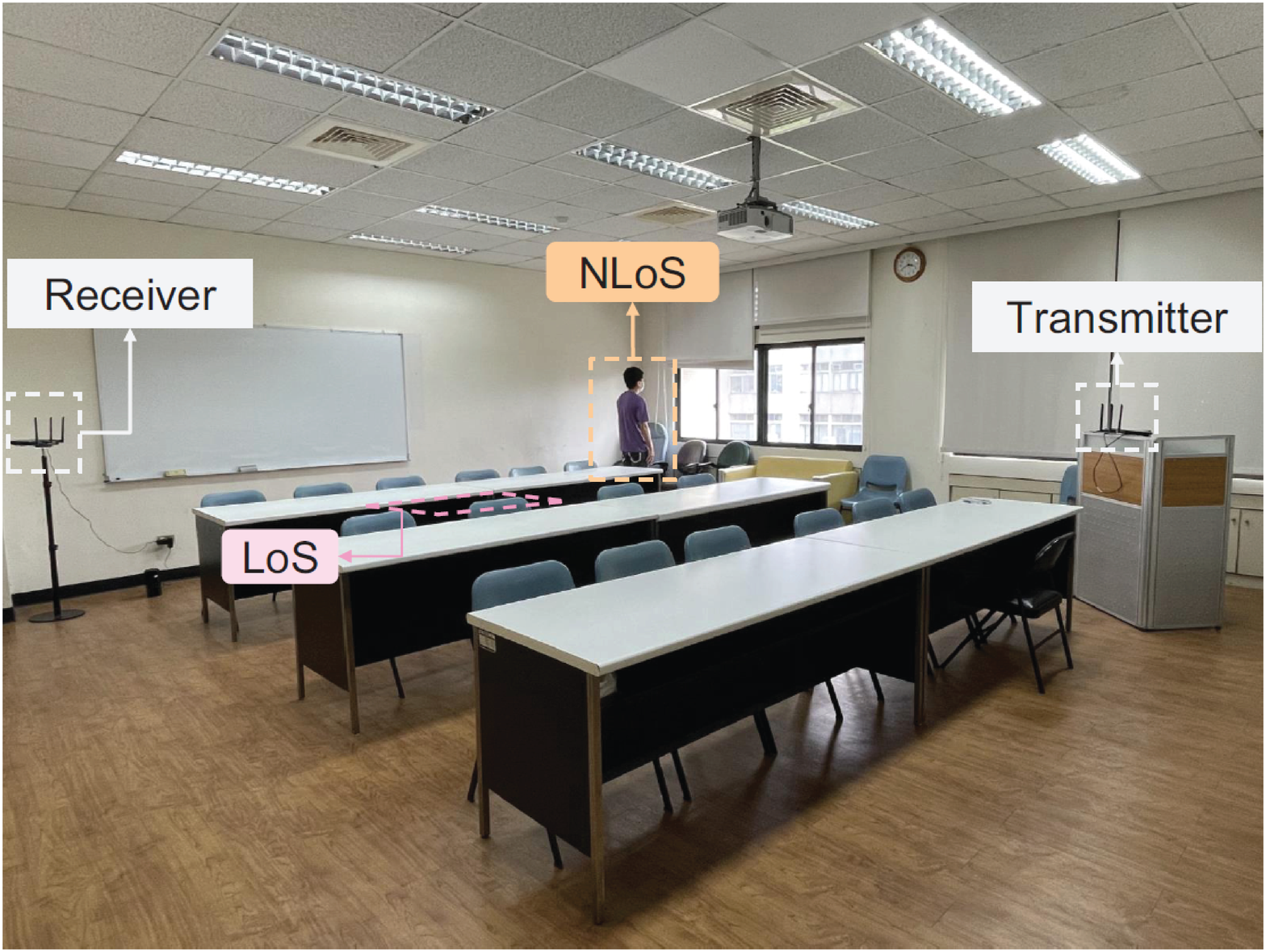}
    \caption{\footnotesize}
    \label{fig:801_real}
  \end{subfigure} 
  \begin{subfigure}[b]{0.20\textwidth}
    \includegraphics[width=\textwidth]{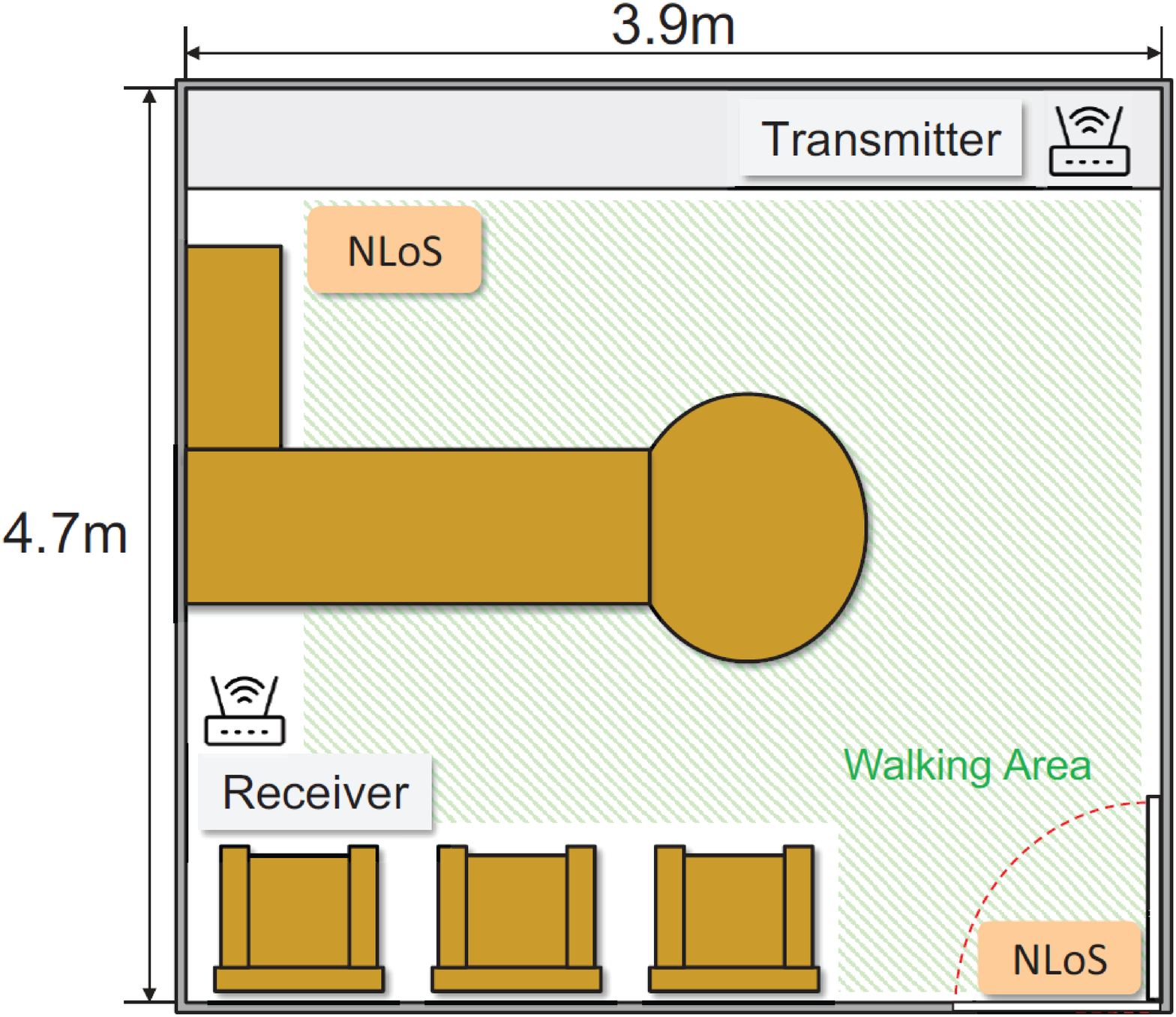}
    \caption{\footnotesize}
    \label{fig:2F_layout}
  \end{subfigure}
  \begin{subfigure}[b]{0.27\textwidth}
    \includegraphics[width=\textwidth]{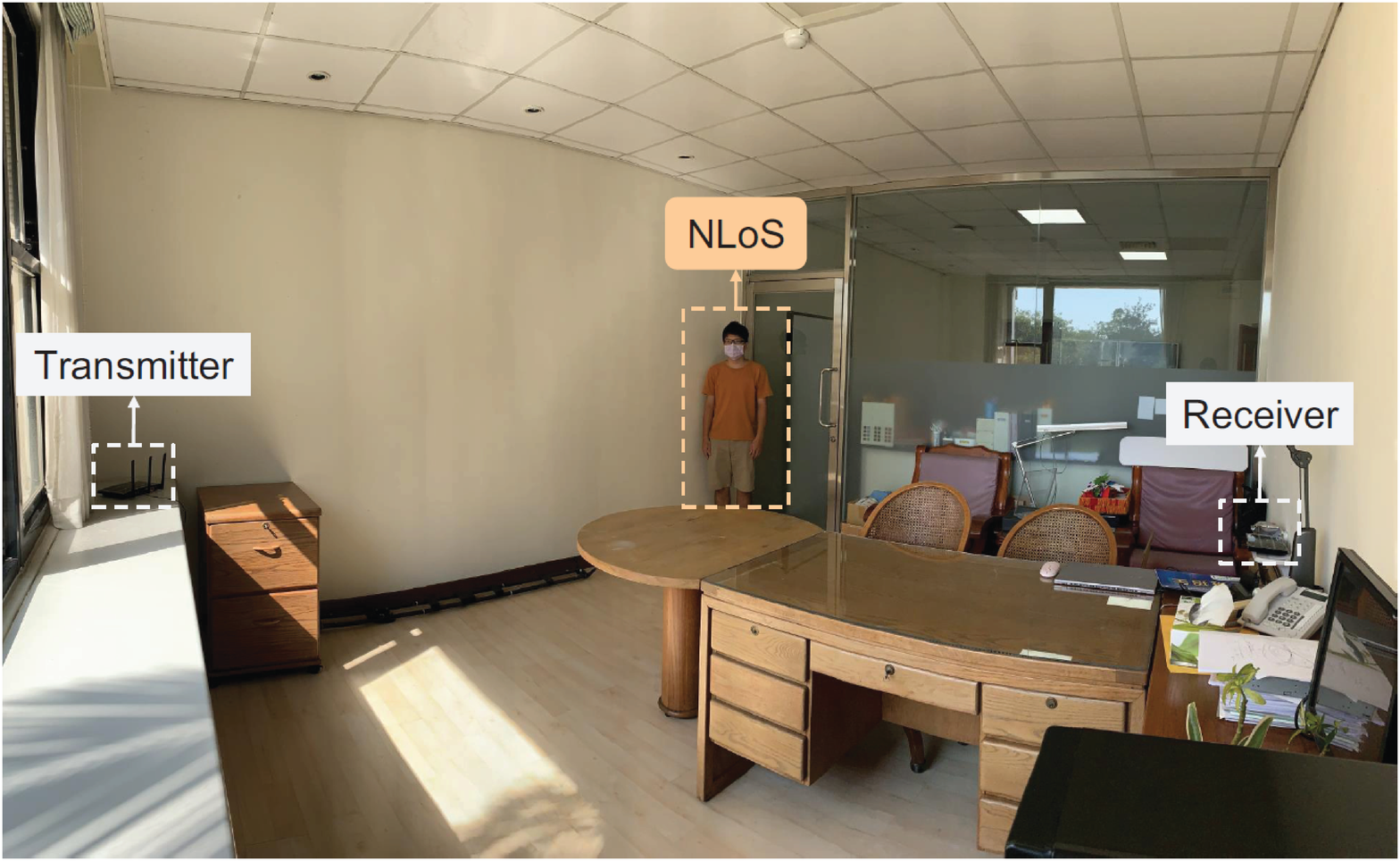}
    \caption{\footnotesize}
    \label{fig:2F_real}
  \end{subfigure}
\caption{The (a) layout and (b) environment of a conference room for scenario 1. The (c) layout and (d) environment of an office for scenario 2.}
\label{exp_scenario}
\end{figure}

\begin{table}[!t]
\footnotesize
\centering
\caption{Interfered/Non-Interfered Detection}
\begin{tabular}{|c|c|c|}
\hline
Case & Interference & Non-Interfered \\ \hline
Averaged Accuracy & 94.18$\%$ & 95.28$\%$ \\ \hline
\end{tabular}
\label{Interf}
\end{table}

\section{Performance Evaluation} \label{CH_Exp}

\subsection{Experimental Setup}
In our experiments, we evaluate the performance of the proposed CRONOS system under two scenarios in the practical environment in Fig. \ref{exp_scenario}. In Figs. \ref{fig:801_layout} and \ref{fig:801_real}, it reveals a conference room with a size of $7 \times 6.75$ $m^2$, whereas a smaller office of $4.7 \times 3.9$ $m^2$ is tested in Figs. \ref{fig:2F_layout} and \ref{fig:2F_real}. The transmitter and receiver are placed at two respective corners, provoking the worst case of static NLoS problem. We describe the detailed setting for each case as follows:
\begin{itemize}
	\item Case 1 (\textbf{Empty}): Data of empty case are collected when no one is in presence in the room.
	\item Case 2 (\textbf{Static NLoS}): When collecting the data of NLoS static case, a person has to stand still at two NLoS corners where no APs are placed. Therefore, it generates some NLoS signal paths where human are present (Orange NLoS area).
	\item Case 3 (\textbf{Static LoS}): The person being tested should stand in line with both the transmitter and receiver, as shown in pink LoS area with dotted link between two APs. Note that this case is only conducted in the first scenario.
	\item Case 4 (\textbf{Dynamic}): The dynamic moving case involves a tester arbitrarily walking around the green area.
\end{itemize}
We further notice that all datasets are collected when a person is inside the room. There does not exist people outside the room during the data collection, which may degrade the performance of presence detection. As shown in Table \ref{Interf}, we evaluate the interference case in scenario 1 with and without a person arbitrarily walking outside the room. The interfered environment has around $1\%$ accuracy degradation compared to the data without human presence interference\textsuperscript{\ref{note1}}\footnotetext[2]{Detection interference highly depends on the materials of walls and windows as well as the number of people and the human behaviour outside the room. Such interference imposes a compelling challenge and is complex to be considered, which can be left as future research works. \label{note1}}. Moreover, the movement case of arbitrary human walking will include both cases of LoS and NLoS movement. We can observe from Fig. \ref{exp_scenario} that most of time the person is walking in the NLoS path owing to arbitrary walking manner.

Two Wi-Fi APs with a central frequency of 2.447 GHz and a bandwidth of 20 MHz are installed in the indoor environment, one as the transmitter and the other as the receiver, with each router having two antennas. This allows us to receive the CSI data of four transmission pairs in one packet, and the maximum number of pairs for CSI ratios is $\tbinom{4}{2}=6$. Furthermore, each transmission pair comprises $K=56$ subcarrier values of the CSI data. Using this setup, data for the four cases will be collected at a sampling rate of 10 Hz. Table \ref{data_sz} presents the data sizes of the four cases for both training and testing data in both experimental fields, whilst the remaining parameters applied in our system are summarized in Table \ref{parameter}.

	We set the RP window size $\tau=50$ to ensure that it contains enough timestamps. The RP thresholds for scenarios 1 and 2 are respectively set as 10 and 33 based on CDF of $\bm{D}_e$ in $\eqref{eq:can_set}$ to be $0.9$. The window size is set as $\tau_{\gamma}=\tau_c=2000$ for scenario 1 and $\tau_{\gamma}=\tau_c=1000$ for scenario 2. The number of transmission pairs is set as $Q=3$ and $Q=2$ in scenarios 1 and 2, respectively, taking into account the room size. The sizes of RPs and coloring CSI ratio images are both fixed as $32 \times 32$. As for data augmentation, we set the scale of random resize cropping to $(0.2, 1)$, and the probability of random horizontal flipping $\varepsilon=0.5$. In training process, the scalar temperature parameter is set as $\zeta=0.07$ for supervised contrastive loss \cite{supcon}. The weight for the consultation loss is empirically set as $\lambda=0.5$. We define $recall = \frac{TP}{TP+FN}$ and $precision = \frac{TP}{TP+FP}$ to evaluate the performance of the classification model, where $TP$ represents true positive, $FP$ indicates false positive, $TN$ stands for true negative, and $FN$ represents false negative. The F1-score for each class can be obtained as
\begin{equation}
F1-score = 2\times \frac{recall\times precision}{recall+precision}.
\end{equation}
To evaluate the overall performance of all cases, we calculate the average F1-score, which is the average of the F1-scores of the four cases. It is important to note that we average the results of ten trials to obtain the final result for each experiment.

\subsection{Effect of Different Input Images}

\begin{figure}
\centering
\includegraphics[width=3.3in]{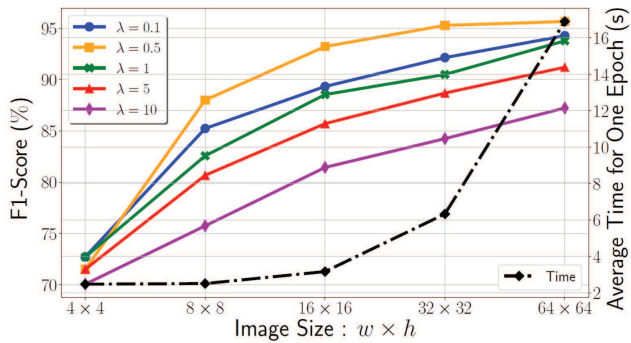}
\caption{The performance of F1-score versus different image sizes and weights of consultation loss $\lambda$.}
\label{fig:size_vs_lambda}
\end{figure}

\begin{figure}
\centering
\includegraphics[width=3.3in]{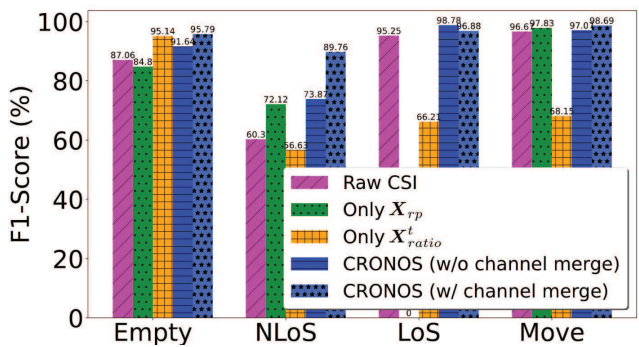}
\caption{The performance comparison of utilization of raw CSI data, only $\bm{X}_{rp}$, only $\bm{X}^t_{ratio}$ as well as proposed CRONOS with and without channel merging mechanism.}
\label{fig:one_input}
\end{figure}

\begin{figure}
\centering
\includegraphics[width=3.3in]{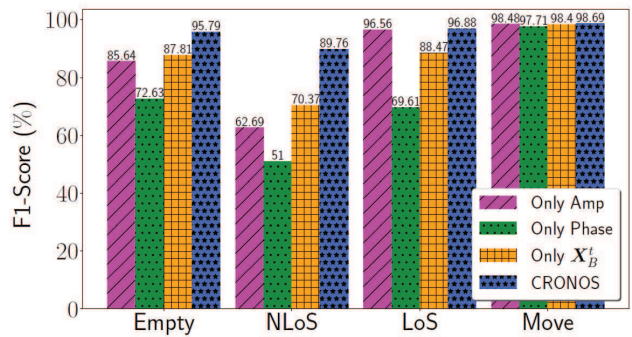}
\caption{The performance comparison of utilization of only CSI amplitude, phase, $\bm{X}^B_{t}$ and proposed CRONOS.}
\label{fig:diff_input}
\end{figure}

\begin{figure*}
  \centering
  \begin{subfigure}[b]{0.3\textwidth}
    \includegraphics[width=\textwidth]{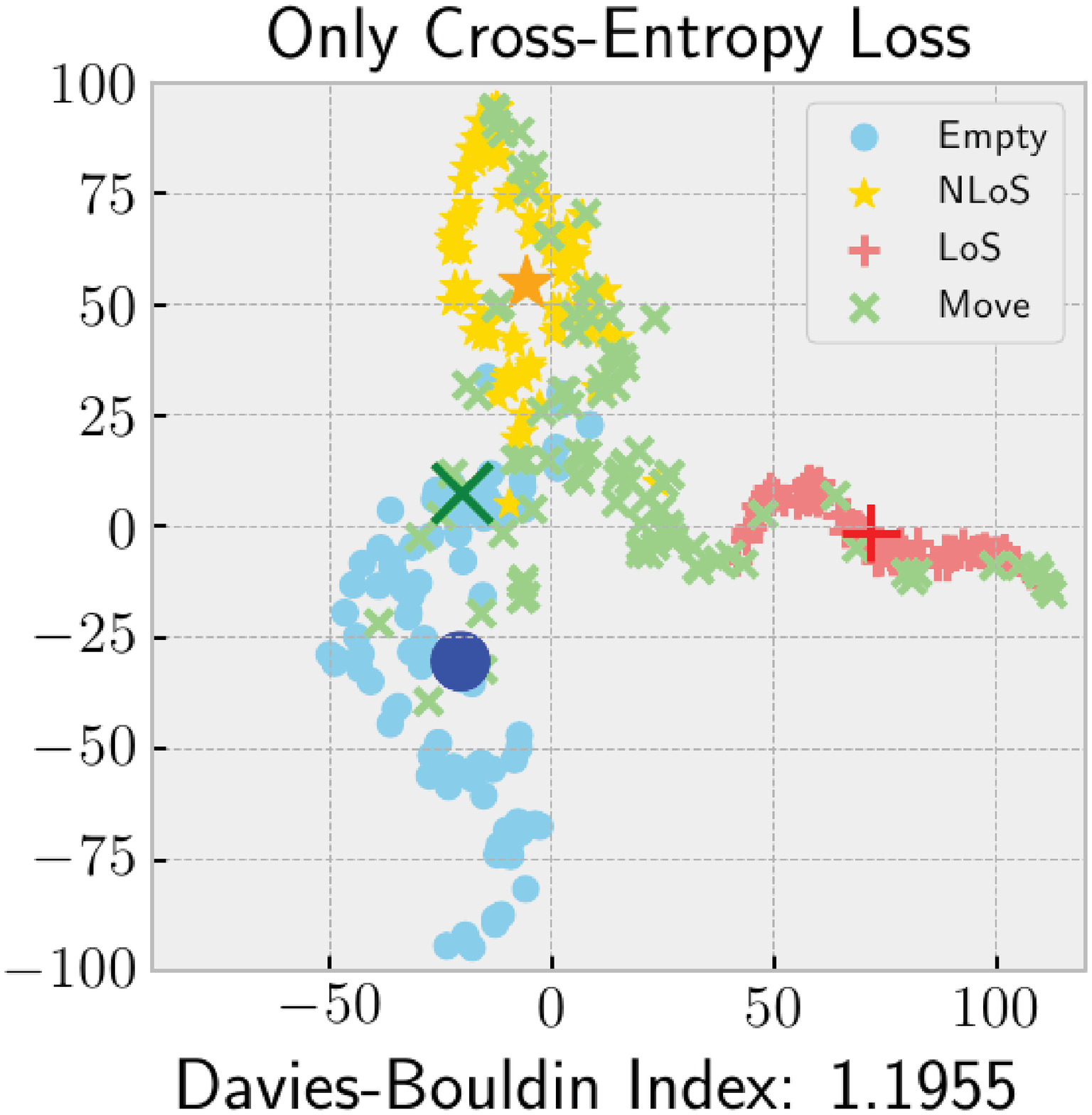}
    \caption{\footnotesize}
    \label{fig:tsne_a}
  \end{subfigure}
  \begin{subfigure}[b]{0.3\textwidth}
    \includegraphics[width=\textwidth]{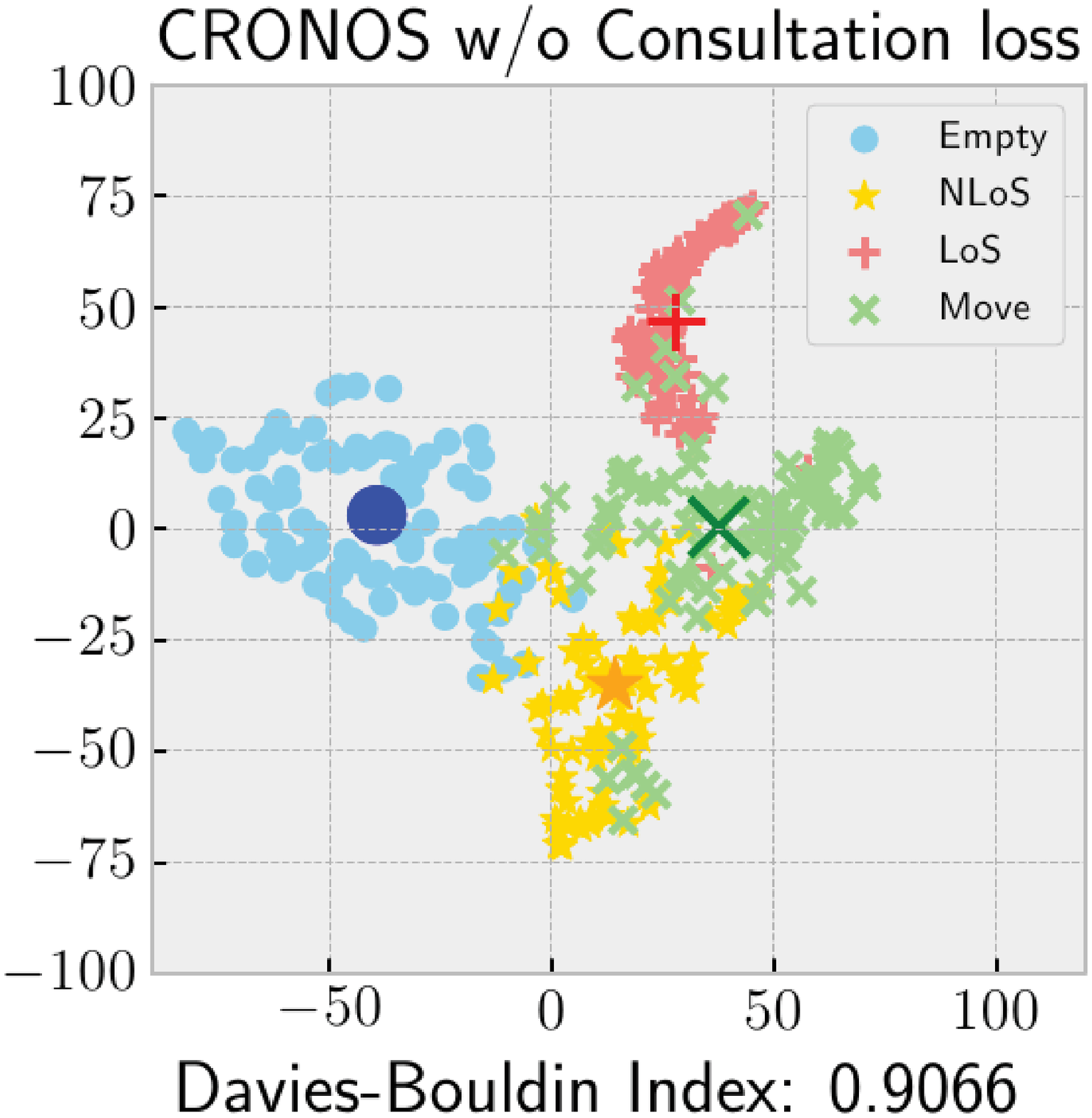}
    \caption{\footnotesize}
    \label{fig:tsne_b}
  \end{subfigure}
  \begin{subfigure}[b]{0.3\textwidth}
    \includegraphics[width=\textwidth]{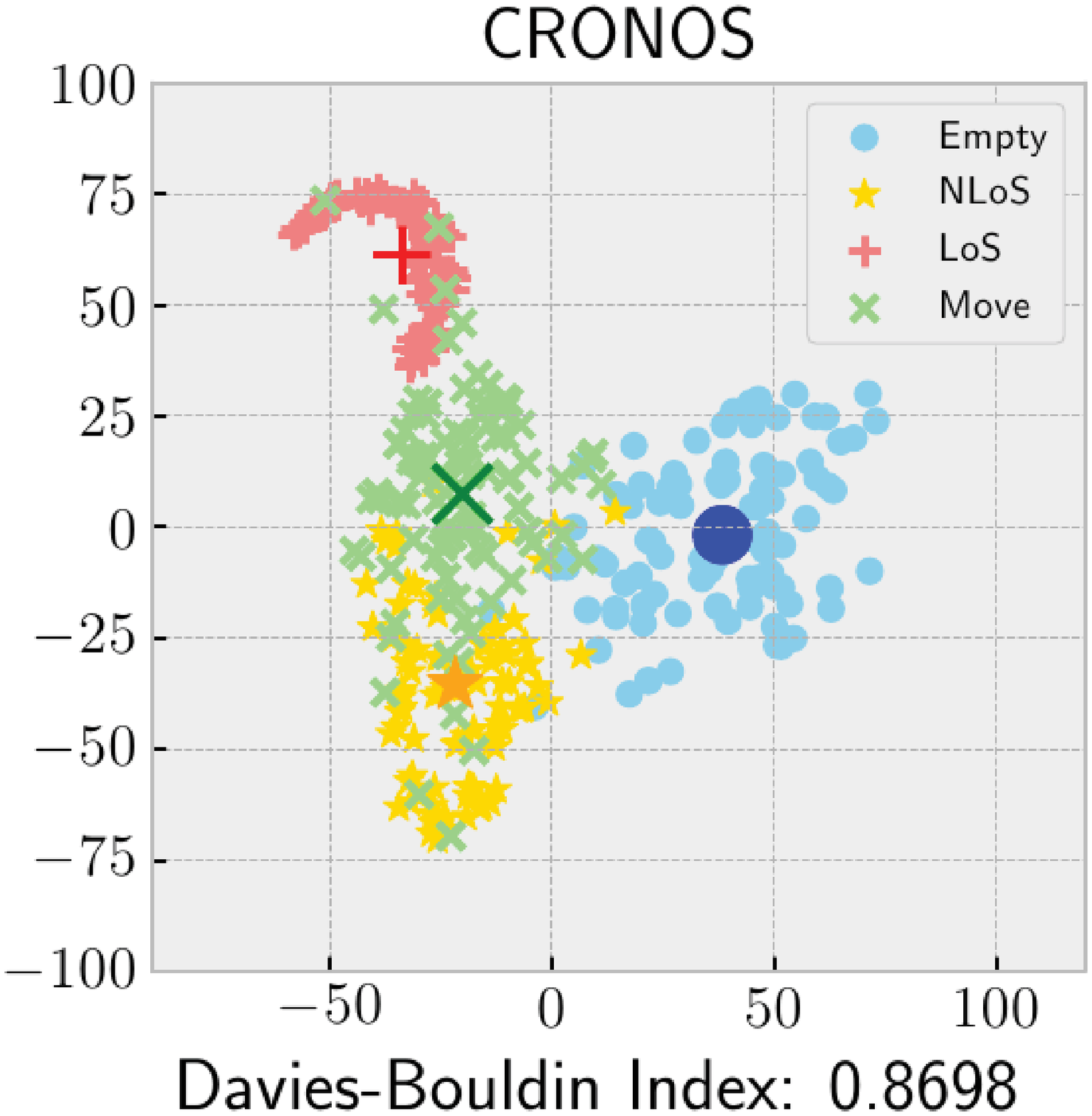}
    \caption{\footnotesize}
    \label{fig:tsne_c}
  \end{subfigure}
\caption{The result of 2D representation employing different loss functions: (a) Only cross-entropy loss (b) CRONOS system without consultation loss (c) Proposed CRONOS system.}
\label{fig:tsne}
\end{figure*}

In \fig\ref{fig:size_vs_lambda}, we compare the effects of different image sizes and $\lambda$ on performance and training time by utilizing the data in scenario 1. We can observe that larger image sizes can provide a higher detection F1-score, as they provide a higher resolution for capturing informative image features. However, if the image size becomes too large, it will significantly increases the computational complexity in training. For example, albeit the F1-score of around $95.2\%$ the training using the image size of $64\times 64$ takes triple time longer than that of $32\times 32$, with asymptotic F1-score. Additionally, we infer that a smaller value of weights for consultation loss does not provide the highest accuracy. With larger $\lambda$ resulting in the consultation loss to dominate the training process, the representations of different cases cannot be effectively separated, leading to a lower accuracy. Therefore, we select a moderate candidate of the image size of $32\times 32$ as well as $\lambda=0.5$ in the following evaluations.

In \fig\ref{fig:one_input}, we compare the impact of different input images on the performance by utilizing the data in scenario 1, i.e., adoption of (1) raw CSI data, (2) only $\bm{X}_{rp}$, (3) only $\bm{X}^t_{ratio}$ (4) CRONOS without channel merging and (5) CRONOS with channel merging. Four cases of human presence detection are evaluated, with their respective F1-scores. Note that using a single type of input image cannot be implemented through consultation loss and S3FEC. As a result, we only employ supervised contrastive loss to train their encoders. As for CRONOS scheme with or without channel merging, they both employ the features of $\bm{X}_{rp}$ and $\bm{X}^t_{ratio}$. In \fig\ref{fig:one_input}, we can observe that using only $\bm{X}_{rp}$ results in poor F1-scores for the first three static cases, especially LoS case. This is because the amplitude difference in these situations is too small to exceed the pre-defined threshold of $\gamma$, causing their RPs to become almost black and leading to classification errors, as shown in \fig\ref{fig:input_image}. Nevertheless, RP is very accurate in classifying the moving case, which achieves an F1-score of 97.83$\%$. Therefore, RP can indeed be applied to differentiate between dynamic and static cases. On the other hand, utilizing only $\bm{X}^t_{ratio}$ results in low F1-scores for NLoS, LoS, and moving cases. The reason for this is that $\bm{X}^t_{ratio}$ only includes information from one packet, leading to ambiguity between human walking and standing. Furthermore, benefited by the channel merging mechanism in CRONOS, NLoS feature can be well extracted among the static cases. Therefore, it is envisioned that the channel merging enhanced CRONOS system with both inputs of $\bm{X}_{rp}$ and $\bm{X}^t_{ratio}$ can achieve high accuracy performance due to complementary impact from the designed consultation loss and S3FEC.

As illustrated in \fig\ref{fig:diff_input}, we replace $\bm{X}^t_{ratio}$ with amplitude, phase, and binary CSI ratio images $\bm{X}^t_{B,q}$ generated from equation $\eqref{eq:binary_image}$ by utilizing the data in scenario 1. Note that the amplitude and phase images are binary, as presented in \fig\ref{fig:input_image}, and we concatenate all four transmission pairs as the input. As shown in \fig\ref{fig:diff_input}, both the scheme with only $\bm{X}^t_{B}$ and CRONOS that adopt CSI ratio can provide higher F1-scores than that with only amplitude and phase images when classifying empty and NLoS cases in most of circumstances. This is because that the CSI ratio eliminates the random phase offset by combining the joint information of amplitude and phase. Accordingly, the CSI ratio is more sensitive than only amplitude or phase information for the stationary person at corner. However, the performance of $\bm{X}^t_{B}$ in LoS is worse than that of amplitude images due to apparent fluctuation when a person blocking the LoS path. In this context, CSI ratio is capable of detecting the NLoS case because it includes both information. Therefore, the difference between empty and NLoS cases becomes observable in the form of complex-plane images. Benefited by both shapes and colorization schemes, the proposed CRONOS system accomplishes the highest F1-score than the other methods using only a single types of information.

\subsection{Effect of Supervised Contrastive Loss and Consultation Loss}

We provide visualizations of representations learned by different loss functions in Fig. \ref{fig:tsne}, including cross-entropy loss, and CRONOS without and with consultation loss for the collected data in scenario 1. The $t$-distributed stochastic neighbor embedding (t-SNE) \cite{tsne} is applied to reduce the 512-dimensional representations into a generic 2D visualizations. As shown in \fig\ref{fig:tsne}, we randomly sample 100 testing data per case with four noticeable marks denote the geometric centers of each class. Additionally, we select the average Davies-Bouldin (DB) index \cite{db_index} of 10 trials to evaluate the cluster validity of these loss functions. A smaller value of DB index indicates a better clustering result, i.e., data within the same cluster gets closer to each other, and data belonging to different clusters becomes farther apart.

We can observe from Fig. \ref{fig:tsne_a} that cross-entropy loss results in highly-overlapped representative features for all cases owing to its consideration of only final prediction. However, as depicted in Fig. \ref{fig:tsne_b}, CRONOS without consultation loss has a comparably lower DB index than that of cross-entropy, implying that employing supervised contrastive learning is capable of separating the hidden representative features for different classes. Furthermore, as shown in Fig. \ref{fig:tsne_c}, it reveals that the CRONOS system achieves the lowest DB index, which means that the designed consultation loss can effectively generate appropriate representation of the CSI ratio for prediction. To elaborate a little further, we calculate the Euclidean distance between the geometric centers of the three static cases in Fig. \ref{fig:tsne}. The distances of CRONOS with consultation loss between cases $(1,2)$, $(1,3)$, and $(2,3)$ are 68.95, 95.79, and 97.4, respectively, whereas those without consultation loss are respectively 65.87, 80.24, and 82.85. Accordingly, it can also be shown that the proposed consultation loss can increase the distance between the representations of the three static cases, which resolving ambiguity between empty room and NLoS static case.

\begin{figure}
\centering
\includegraphics[width=3.3in]{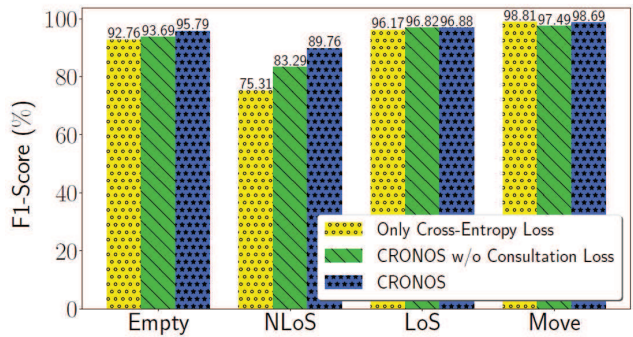}
\caption{The performance comparison between different loss combinations.}
\label{fig:diff_loss}
\end{figure}

Moreover, as shown in \fig\ref{fig:diff_loss}, it is evident that both schemes with supervised contrastive learning outperform that with only cross-entropy loss. This is because the model trained with only cross-entropy loss extracts features solely from the input images, which leads to an inadequate representation, as explained previously in \fig\ref{fig:tsne_a}. By contrast, supervised contrastive learning separates the four cases by contrastive loss, which generates appropriate representations for classification compared to the scheme using only cross-entropy loss. Compared to the scheme without consultation loss, the CRONOS system with the design of consultation loss improves the F1-score by 2.1$\%$, 6.47$\%$, and 0.06$\%$ in cases 1, 2, and 3, respectively. The significant gain on F1-score for case 2 of NLoS static scenario can be observed.

\par

\subsection{Effect of Different Model Architectures}

\begin{figure}
  \centering
  \begin{subfigure}[b]{0.47\textwidth}
    \includegraphics[width=\textwidth]{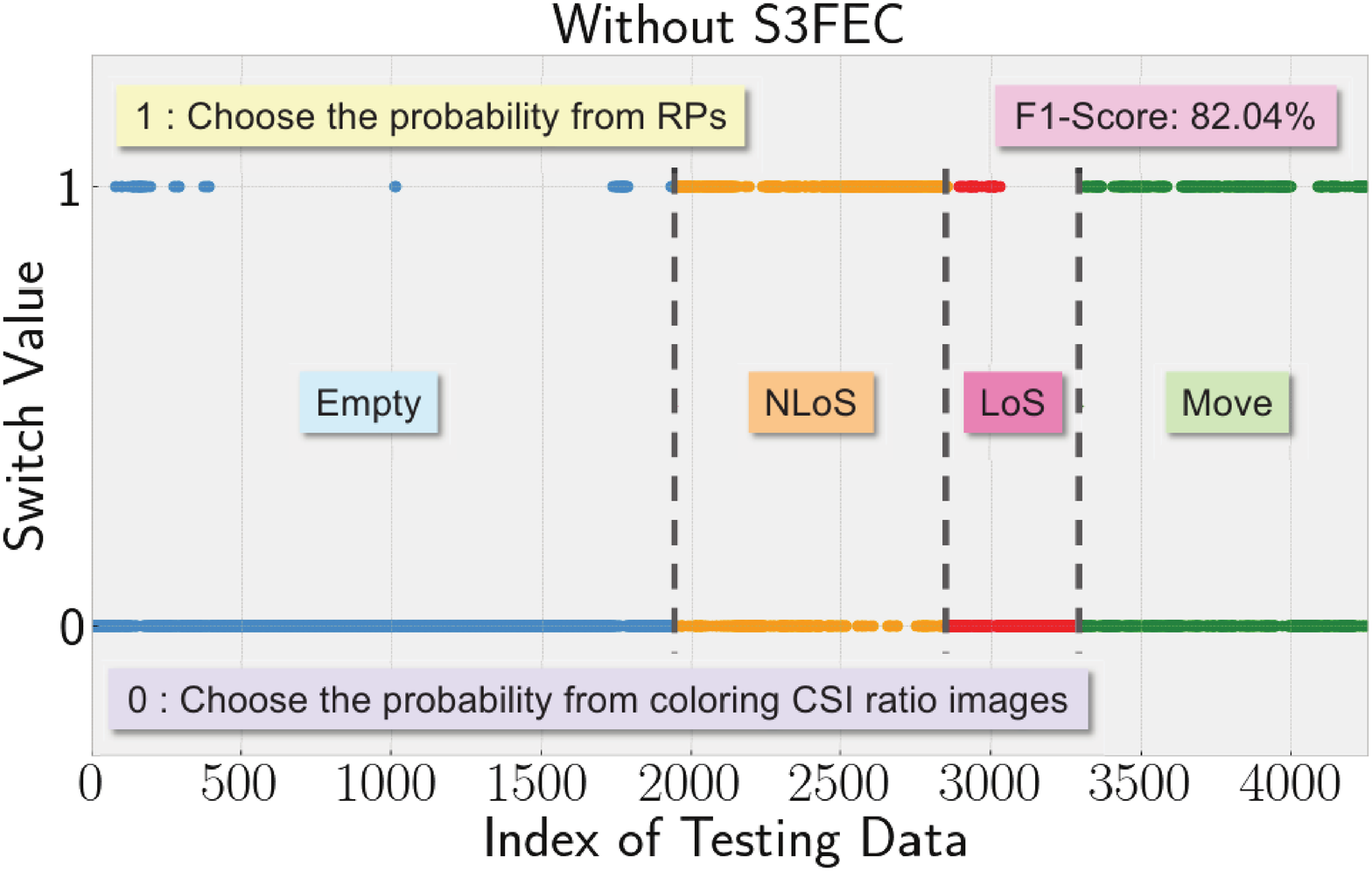}
    \caption{\footnotesize}
    \label{fig:sv_wo_s3fec}
  \end{subfigure}
  \begin{subfigure}[b]{0.47\textwidth}
    \includegraphics[width=\textwidth]{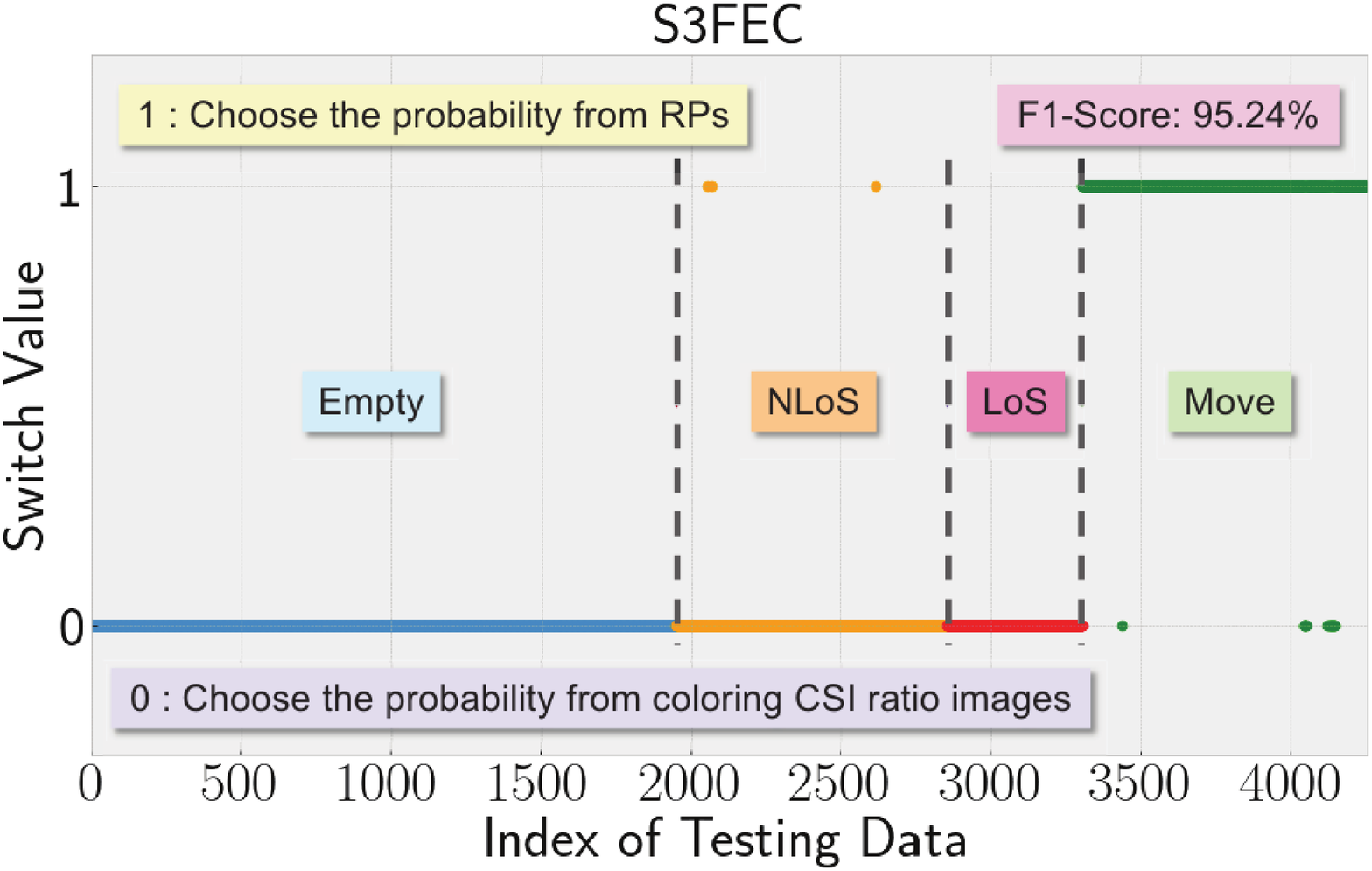}
    \caption{\footnotesize}
    \label{fig:sv_s3fec}
  \end{subfigure}
\caption{Switch values comparison between (a) without S3FEC and (b) with S3FEC.}
\label{fig:switch_value}
\end{figure}

In \fig\ref{fig:switch_value}, we present the switch values in CRONOS with and without S3FEC. The horizontal axis represents the index of testing data in scenario 1, while the vertical axis is the switch values. Note that $\omega$ is either 0 or 1, i.e., $\omega=1$ means that the model uses $\hat{\bm{y}}_{d}$ as the final classification probability, while $\omega=0$ denotes that $\hat{\bm{y}}_{s}$ is utilized for prediction. As depicted in \fig\ref{fig:sv_wo_s3fec}, the model without S3FEC tends to rely more on the probabilities from RPs to classify the three static cases. However, as the RPs of the three static cases become similar, they will be misclassified from each other. Moreover, when classifying moving situation without S3FEC, CSI ratio will be employed for prediction, leading to  a lower F1-score of 82.04$\%$ compared to that with S3FEC. However, as depicted in \fig\ref{fig:sv_s3fec}, the proposed S3FEC scheme is capable of dynamically selecting either RPs or CSI ratios to classify all cases, which achieves a higher F1-score of 95.24$\%$.

\begin{figure}
\centering
\includegraphics[width=3.3in]{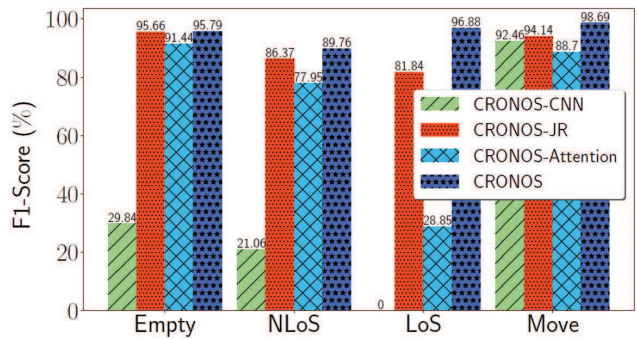}
\caption{The performance comparison between different model architectures.}
\label{fig:diff_model}
\end{figure}

Here, we study the impact of different model architectures on the performance of our system. Specifically, we will evaluate three alternatives to our original architecture. The first alternative replaces the ResNet-18 of stages 1 and 2 with a two-layer CNN architecture. The second alternative replaces S3FEC with the JR method, which concatenates the two representation vectors with a linear classifier. Finally, we will test an attention-based approach, where we multiply $\hat{\bm{y}}_{d}$ and $\hat{\bm{y}}_{ratio}$ with model weights coming from their respective representations in an FCN. As shown in \fig\ref{fig:diff_model}, CRONOS-CNN has the worst F1-score since loss functions in stages 1 and 2 cannot be optimized from a single overfitting neural network. As a result, attention-based network is adopted to provide a more complex architecture. It was expected that the specific feature for each class will be more focused on. However, it has a moderate F1-score owing to the unawareness of model types where it assigns attention. Alternatively, CRONOS-JR can achieve higher performance since RPs are effective at classifying dynamic and static cases, whilst coloring CSI ratios are effective at distinguishing between three static situations. Nonetheless, as explained in section \ref{stage3}, JR directly concatenating representations constrains cases among each other. In contrast, the proposed CRONOS with S3FEC performs flexible selection between RPs and CSI ratio, which has the highest F1-score than existing methods.

\begin{figure}
\centering
\includegraphics[width=3.3in]{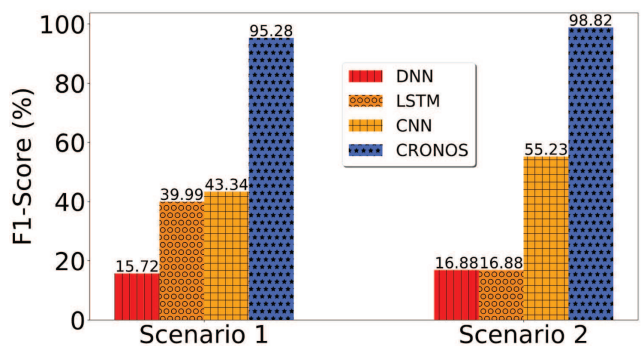}
\caption{The performance comparison of proposed CRONOS system with the baselines of DNN, CNN and LSTM in scenarios 1 and 2.}
\label{baseline_nn}
\end{figure}

\subsection{Benchmark Comparison}

As depicted Fig. \ref{baseline_nn}, we firstly compare the performance of our proposed CRONOS with the existing baselines of deep neural network (DNN), CNN and LSTM in both scenarios 1 and 2. For a fair comparison, we employ the processed data associated with grey images and channel merging. Note that all three baselines have the same images as input as CRONOS for classification. We can observe that the primitive neural network with simple structures in DNN and CNN cannot extract the CSI picture having small changes in either amplitude or phase difference. Furthermore, even though LSTM is beneficial to the time-correlated information, it inappropriate to distinguish the image-based features. On the other hand, our proposed CRONOS system performs the three-stage supervised contrastive learning, achieving the highest F1-score amongst the baselines.

Moreover, we compare the performance of our proposed CRONOS in two experimental scenarios with the existing benchmarks in open literature, which are elaborated as follows.
\begin{itemize}
	\item \textbf{LCDNN} \cite{lcdnn} is designed based on a local connection based deep neural network to extract the features of adjacent subcarriers. It generates a position-dependent local feature based on CSI amplitude for indoor localization. 
	
	\item \textbf{F-LSTM} \cite{f_lstm} first utilizes a Butterworth low-pass filter on the CSI amplitude to suppress noise, then uses a moving average filter to reduce the influence of out-of-range amplitude values and burst noises. Afterwards, LSTM model is employed to classify these filtered CSI amplitudes for human presence detection.

	\item \textbf{P-CNN} \cite{p_cnn} first also performs preprocessing on CSI amplitude and phase information and converts them into images. Two parallel CNNs are employed for these two images for human presence detection. 
	
	\item \textbf{C-MuRP} \cite{fangyu} processes the CSI amplitude and inputs it into a conditional gated recurrent unit (GRU) for detection.
	
	\item \textbf{CALPD} \cite{calpd} employs CSI ratio images and a CNN model for human presence detection. 
\end{itemize}

\begin{table*}[!ht]
\centering
\caption{Performance Comparison in Scenario 1}
\footnotesize
\begin{tabular}{|c|c|c|c|c|c|c|c|c|c|c|c|c|c|c|c|}
\hline
 & \multicolumn{3}{c|}{Empty Room} & \multicolumn{3}{c|}{NLoS static case} & \multicolumn{3}{c|}{LoS static case} & \multicolumn{3}{c|}{Moving case} & \multicolumn{3}{c|}{Average} \\ \hline
Approach & Recall & F1 & $\Delta$F1 & Recall & F1 & $\Delta$F1 & Recall & F1 & $\Delta$F1 & Recall & F1 & $\Delta$F1 & Recall & F1 & $\Delta$F1\\ \hline
LCDNN \cite{lcdnn}	&	91.38	&	83.6	&	12.19	&	48.54	&	56.29	&	33.47	&	84.14	&	84.29	&	12.59	&	80.12	&	79.33	&	19.36	&	76.04	&	75.88	&	19.4 \\
F-LSTM \cite{f_lstm}	&	90.74	&	85.31	&	10.48	&	44.45	&	50.55	&	39.21	&	88.23	&	89.43	&	7.45	&	91.01	&	89.35	&	9.34	&	78.61	&	78.66	&	16.62\\
P-CNN \cite{p_cnn}	&	49.56	&	55.77	&	40.02	&	38.51	&	26.62	&	63.14	&	35.26	&	51.81	&	45.07	&	\textbf{99.98}	&	\textbf{99.99}	&	-1.3	&	55.83	&	58.55	&	36.73 \\
C-MuRP \cite{fangyu}	&	97.43	&	95.52	&	0.27	&	57.56	&	71.13	&	18.63	&	96.56	&	\textbf{97.5}	&	-0.62	&	97.75	&	84.34	&	14.35	&	87.32	&	87.12	&	8.16 \\
CALPD \cite{calpd}	&	80.7	&	82.16	&	13.63	&	66.08	&	65.55	&	24.21	&	77.08	&	80.55	&	16.33	&	84.92	&	79.48	&	19.21	&	77.19	&	76.94	&	18.34 \\ \hline
w/o $\bm{X}^t_{ratio}$	&	92.14	&	84.8	&	10.99	&	78.14	&	72.12	&	17.64	&	0	&	0	&	96.88	&	95.99	&	97.83	&	0.86	&	66.57	&	63.69	&	31.59 \\
w/o $\bm{X}_{rp}$	&	95.44	&	95.14	&	0.65	&	42.41	&	56.63	&	33.13	&	50.71	&	66.21	&	30.67	&	91.62	&	68.15	&	30.54	&	70.04	&	71.53	&	23.75 \\
w/o $L_{sc}$	&	96.6	&	92.76	&	3.03	&	71.18	&	75.31	&	14.45	&	\textbf{99.93}	&	96.17	&	0.71	&	98.29	&	98.81	&	-0.12	&	91.5	&	90.76	&	4.52 \\
w/o $L_{cs}$	&	98.15	&	93.69	&	2.1	&	75.8	&	83.29	&	6.47	&	94.9	&	96.82	&	0.06	&	97.38	&	97.49	&	1.2	&	91.56	&	92.82	&	2.46 \\
w/o S3FEC	&	\textbf{98.9}	&	95.66	&	0.13	&	81.43	&	86.37	&	3.39	&	70.98	&	81.84	&	15.04	&	98.88	&	94.14	&	4.55	&	87.55	&	89.5	&	5.78 \\ \hline
CRONOS	&	97.96	&	\textbf{95.79}	&	-	&	\textbf{85.93}	&	\textbf{89.76}	&	-	&	96.79	&	96.88	&	-	&	98.16	&	98.69	&	-	&	\textbf{94.71}	&	\textbf{95.28}	&	- \\ \hline
 
\end{tabular}
\label{OP1}
\end{table*}

\begin{figure}
\centering
\includegraphics[width=3.3in]{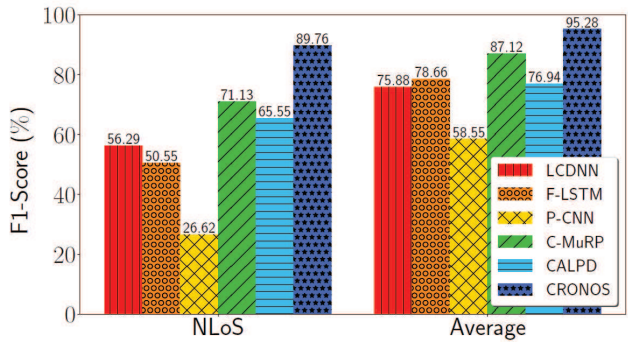}
\caption{The performance comparison of proposed CRONOS system with the existing benchmarks in scenario 1. The NLoS case and the average of all cases are considered.}
\label{fig:baseline_801}
\end{figure}

As presented in Table \ref{OP1}, we compare the recall and F1-score of CRONOS with other existing methods in scenario 1. Note that $\Delta$F1 represents the difference with the F1-score of CRONOS. The schemes with a term of "w/o" indicates the sub-schemes of CRONOS without those factors for ablation studies. We extract the F1-scores of the NLoS case and average results in \fig\ref{fig:baseline_801}. The results reveal that CRONOS outperforms the benchmarks in classifying empty and NLoS static cases, i.e., CRONOS improves recall by 0.53$\%$ and F1-score by 0.27$\%$ compared to the best benchmark of C-MuRP in empty case. While, in NLoS case, CRONOS yields a 19.85$\%$ improvement in recall and 18.63$\%$ in F1-score compared to the best benchmarks of CALPD and C-MuRP, respectively. Although CRONOS may not provide the best performance in LoS and moving cases, it can achieve asymptotic performances to those benchmarks. Therefore, as depicted in \fig\ref{fig:baseline_801}, our CRONOS system outperforms all benchmarks on average performance, indicating that our system can classify all cases of human presence detection.

Furthermore, as presented in the lower half of Table \ref{OP1}, we have conducted ablation studies to verify the validity of CRONOS. It can be seen that the lack of any strategy will lead to a declined F1-score in most cases, e.g., removing either $\bm{X}_{rp}$ or $\bm{X}^t_{ratio}$ severely degrades performance due to unclassifiable features in respective cases. Additionally, supervised contrastive learning in required to train ambiguous representations with contrastive loss. The proposed consultation loss capable of separating representations escalates F1-scores especially for the three static cases, i.e., 6.47$\%$ improvement in NLoS case. To elaborate further, the proposed S3FEC mechanism enables RPs and CSI ratio to perform their respective benefits in classification, resulting in better performance. As a result, CRONOS incorporating all designs achieves the highest recall of 94.71$\%$ and F1-score of 95.28$\%$ in human presence detection.

\par

\begin{table*}[!ht]
\centering
\caption{Performance Comparison in Scenario 2}
\footnotesize
\begin{tabular}{|c|c|c|c|c|c|c|c|c|c|c|c|c|}
\hline
 & \multicolumn{3}{c|}{Empty Room} & \multicolumn{3}{c|}{NLoS static case} & \multicolumn{3}{c|}{Moving case} & \multicolumn{3}{c|}{Average} \\ \hline
Approach & Recall & F1 & $\Delta$F1 & Recall & F1 & $\Delta$F1 & Recall & F1 & $\Delta$F1 & Recall & F1 & $\Delta$F1 \\ \hline
LC-DNN \cite{lcdnn}	&	77.64	&	77.9	&	20.86	&	43.02	&	47.69	&	50.54	&	52.98	&	46.64	&	52.83	&	57.88	&	57.41	&	41.41 \\
F-LSTM \cite{f_lstm}	&	96.91	&	91.91	&	6.85	&	34.17	&	47.21	&	51.02	&	90.41	&	82.04	&	17.43	&	73.83	&	73.72	&	25.1 \\
P-CNN \cite{p_cnn}	&	92.89	&	90.47	&	8.29	&	85.43	&	88.17	&	10.06	&	\textbf{100}	&	\textbf{100}	&	-0.53	&	92.77	&	92.88	&	5.94 \\
C-MuRP \cite{fangyu}	&	97.51	&	91.77	&	6.99	&	25.05	&	39	&	59.23	&	97.35	&	74.62	&	24.85	&	73.3	&	68.47	&	30.35 \\
CALPD \cite{calpd}	&	95.72	&	93.1	&	5.66	&	89.54	&	89.87	&	8.36	&	82.44	&	84.49	&	14.98	&	89.23	&	89.15	&	9.67 \\ \hline
w/o $\bm{X}^t_{ratio}$	&	67.29	&	61.99	&	36.77	&	47.57	&	50.58	&	47.65	&	99.56	&	99.66	&	-0.19	&	71.47	&	70.74	&	28.08 \\
w/o $\bm{X}_{rp}$	&	94	&	93.34	&	5.42	&	65.31	&	73.73	&	24.5	&	93.57	&	82.74	&	16.73	&	84.29	&	83.27	&	15.55 \\
w/o $L_{sc}$	&	99.12	&	96.48	&	2.28	&	89.52	&	93.68	&	4.55	&	\textbf{100}	&	98.29	&	1.18	&	96.21	&	96.15	&	2.67 \\
w/o $L_{cs}$	&	\textbf{99.82}	&	98.23	&	0.53	&	95.63	&	97.51	&	0.72	&	99.61	&	99.4	&	0.07	&	98.36	&	98.38	&	0.44 \\
w/o S3FEC	&	99.81	&	98.57	&	0.19	&	94.8	&	96.89	&	1.34	&	99.04	&	98.26	&	1.21	&	97.89	&	97.91	&	0.91 \\ \hline
CRONOS	&	99.79	&	\textbf{98.76}	&	-	&	\textbf{97.17}	&	\textbf{98.23}	&	-	&	99.44	&	99.47	&	-	&	\textbf{98.8}	&	\textbf{98.82}	&	- \\ \hline
 
\end{tabular}

\label{OP2}
\end{table*}

\begin{figure}
\centering
\includegraphics[width=3.3in]{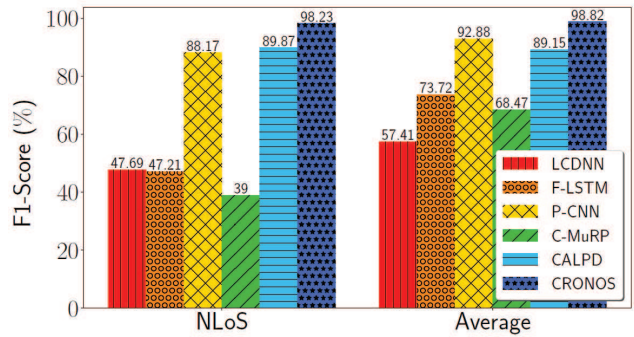}
\caption{The performance comparison of proposed CRONOS system with the existing benchmarks in scenario 2. The NLoS case and the average of all cases are considered.}
\label{fig:baseline_2f}
\end{figure}

In scenario 2, which is conducted in a smaller office room, we have evaluated the performance of our proposed CRONOS system. The results in Table \ref{OP2} and \fig\ref{fig:baseline_2f} demonstrate that CRONOS outperforms all benchmark methods in the NLoS case and average performance. Specifically, in the NLoS case, our system achieves 7.63$\%$ higher recall and 8.36$\%$ higher F1-score than the best benchmark of CALPD. This indicates that CRONOS is capable of effectively alleviating the NLoS static issue even in a smaller scenario. Furthermore, CRONOS outperforms the P-CNN by 6.03$\%$ and 5.94$\%$ in terms of best average recall and F1-score, respectively. Similar to scenario 1, the ablation studies in scenario 2 also envision that all designs are required.

\begin{table}[!t]
\footnotesize
\centering
\caption{Complexity Analysis}
\begin{tabular}{|c|c|c|c|c|}
\hline
Approach & Parameters & \tabincell{c}{Memory \\ Size} & \tabincell{c}{Training \\Time} & \tabincell{c}{Testing\\ Time}\\ \hline \hline
LC-DNN \cite{lcdnn} & 38.33 K & 1.57 MB & 1.2829 s & 2.685 ms \\ \hline
F-LSTM \cite{f_lstm} & 72.34 K & 366.91 MB & 0.629 s & 6.086 ms \\ \hline
P-CNN \cite{p_cnn} & 54.24 K & 85.51 MB & 8.6822 s & 5.933 ms \\ \hline
C-MuRP \cite{fangyu} & 743.11 K & 487.42 MB & 3.011 s & 286.73 ms\\ \hline
CALPD \cite{calpd} & 35.2 K & 68.81 MB & 0.1508 s & 0.228 ms \\ \hline
CRONOS & 22.99 M & 88.56 MB & 6.33 s & 22.212 ms\\ \hline
\end{tabular}
\label{complexity}
\end{table}

At last, we present the analysis of computational complexity of the proposed CRONOS compared to the benchmarks in terms of required total number of deep learning model parameters, required memory size per epoch, and training/testing time. Note that training time is calculated per epoch, whilst testing time is averaged over each batch. We evaluate the system with the 11th Gen Intel(R) Core(TM) i7-11700 CPU and NVIDIA GeForce RTX 3070 Ti GPU. As shown in Table \ref{complexity}, our CRONOS system requires a total of 22.99 M parameters to be trained owing to the complex models reckoning with the respective issues in human presence detection. An acceptable memory size of 88.56 MB is required for CRONOS compared to F-LSTM and C-MuRP methods. The training process of CRONOS takes the moderate time of 6.33 seconds per epoch, whereas it takes around 22 ms of testing time for each batch owing to the 3-stage process. To conclude, the proposed CRONOS system is feasible and implementable in practical applications, which outperforms the other existing solutions in the open literature.

\section{Conclusion} \label{CH_conclusion}
In this paper, we have conceived a device-free human presence detection system that utilizes Wi-Fi CSI to detect a human standing still in NLoS areas. Our proposed CRONOS system includes FEIG and three-stage supervised contrastive learning. FEIG generates RPs from CSI amplitude differences to distinguish dynamic from static situations, whereas CSI ratio images aims for classifying ambiguous static cases. Colorization is further designed to enhance the visual difference between empty and NLoS cases. The designed contrastive learning with consultation loss effectively separates the representations in static cases. Moreover, S3FEC allows the model to automatically select the probability outcome from either RP or coloring CSI ratio. Experimental evaluations reveal that CRONOS achieves the highest average F1-score of 95.28$\%$, resulting in around 18.63$\%$ improvement of F1-score compared to the best benchmark in NLoS case.

\footnotesize
\bibliographystyle{IEEEtran}
\bibliography{IEEEabrv,myReference}

\end{document}